\documentclass[12pt]{article}

\usepackage{threeparttable}
\usepackage{comment}
\usepackage{graphicx}
\usepackage[utf8]{inputenc}
\usepackage{lmodern}
\usepackage[english]{babel}
\usepackage[T1]{fontenc}
\usepackage{amsmath}
\usepackage[dvipsnames]{xcolor}
\usepackage{multirow}
\usepackage{float}
\usepackage{pdflscape}
\usepackage{amssymb}
\usepackage{dcolumn}
\usepackage{caption}
\usepackage{subcaption}

\newcolumntype{d}[1]{D{.}{.}{#1}}

\usepackage[a4paper, top=2.5cm, bottom=2.5cm, right=2.5cm, left=2.5cm]{geometry}
\usepackage[onehalfspacing]{setspace}
\usepackage{booktabs}

\usepackage[authoryear,round]{natbib}
\begin{document}

\setlength{\parindent}{2em} 

\title{During and after COVID-19: What happened to the home advantage in Germany's first football division? \vspace{0.3cm}}

\begin{onehalfspacing}

\author{Thorsten Schank%
\thanks{Johannes Gutenberg-University Mainz (JGU Mainz), Institute of Labor Economics (IZA), and Labor and Socio-Economic Research Center of the University of Erlangen-Nuremberg (LASER). Address: JGU Mainz, Jakob-Welder-Weg 4, 55128 Mainz, \texttt{schank@uni-mainz.de}.}\vspace{-0.3cm}
\\
\small{\emph {JGU Mainz, IZA, LASER}}
\and
Vivien Voigt%
\thanks{Johannes Gutenberg-University Mainz (JGU Mainz). Address: JGU Mainz, Jakob-Welder-Weg 4, 55128 Mainz,\texttt{vivien.voigt@uni-mainz.de}.}\vspace{-0.3cm}
\\
\small{\emph {JGU Mainz}}
\and
Christian Orthey%
\thanks{Johannes Gutenberg-University Mainz (JGU Mainz). Address: JGU Mainz, Jakob-Welder-Weg 4, 55128 Mainz,\texttt{christian.orthey.co@gmail.com}.
\newline
We gratefully acknowledge comments from Martyn Andrews, Mario Bossler, Viktor Bozhinov, Manuel Denzer, and Stefan Schwarz as well as those received at a seminar at the JGU Mainz.}\vspace{-0.3cm}
\\
\small{\emph {JGU Mainz}} \vspace{0.3cm}
}

\date{November 2024}
\maketitle

\vspace{-0.5cm}
\begin{abstract}
\noindent It is well-established that the home advantage (HA), the phenomenon that on average the local team performs better than the visiting team, exists in many sports. In response to the COVID-19 outbreak, spectators were banned from football stadiums, which we leverage as a natural experiment to examine the impact of stadium spectators on HA. Using data from the first division of the German Bundesliga for seasons 2016/17 to 2023/24, we are the first to focus on a longer time horizon and consider not only the first but all three seasons subject to spectator regulations as well as two subsequent seasons without. We confirm previous studies regarding the disappearance of the HA in the last nine matches of season 2019/20. This drop materialised almost entirely through a reduction of home goals. The HA in season 2020/21 (with spectator ban during most matches) was very close to the pre-COVID-19 season 2018/19, indicating that teams became accustomed to the absence of spectators. For season 2021/22, with varying spectator regulations, we detect a U-shaped relationship between HA and the stadium utilisation rate, where HA increases considerably for matches with medium stadium utilisation which is associated with a larger difference in running distance between the home and away teams. 

\vspace{0.4cm}
\phantom{\emph{Keywords}:} football, soccer, Germany
\end{abstract}

\end{onehalfspacing}

\thispagestyle{empty}
\clearpage
\setcounter{page}{1}

\setcounter{page}{1}
\section{Introduction}
The home advantage (HA) in football and other sports is well-established, see \cite{pollard-2017} or \cite{gomez-ruano-et-al-2021} for a comprehensive overview.\footnote{By `football' we refer to the sport that is typically known as `soccer' in the United States and Canada, and as `football' in the rest of the world.} HA denotes the phenomenon that on average the home team performs better than the visiting team. Hence, in football the HA occurs if \textit{ceteris paribus} the (expected) value of the difference between the goals scored by the home team and the goals scored by the away team is positive.\footnote{Analogously, the HA arises if the (expected) probability that the home team wins exceeds the (expected) probability that the home team looses. Both probabilities do not add up to one if draws are possible.} The HA is attributed to several possible causes such as the audience influencing the behaviour and motivation of players, the audience influencing the referee decisions, familiarity of the home team with the field and with specific rules, travel efforts and fatigue of the away team, territoriality, i.e. the home team's response to the away team's invasion of their territory, and psychological effects of expectations differing between both teams (\citeauthor{pollard-2008}, \citeyear{pollard-2008}). 

In spring 2020, the COVID-19 outbreak resulted in a ban of spectators during subsequent matches. Several studies have taken advantage of this quasi-experimental situation to examine the effect of spectators on the HA. \cite{leitner-et-al-2023} document in a systematic review that the majority of studies find that ghost games reduced the HA in football significantly.\footnote{It might be possible, however, that besides the spectator ban, other changes after the start of the pandemic led to the observed match outcomes. More specifically, (i) players might have played with less physical contact to avoid contagion or due to psychological reasons emerging form prolonged social distancing; (ii) there was a break for several weeks after the COVID-19 outbreak before matches resumed, interrupting routines and shifting matches to unusual time of the year; (iii) public pressure might have been different afterwards; (iv) the procedures before the game might have changed; (v) short-term changes in team composition due to positive COVID-19 condition of players.} Likewise, \cite{wang-qin-2023} detect in their literature review a tendency towards a reduced HA. However, they emphasise that reported effects vary by country and are especially strong for the German Bundesliga.

\begin{figure}[ht!]\centering
\caption{Average stadium utilisation across matchdays and seasons in the German Bundesliga}
\label{fig:fig1}
\includegraphics[width=0.7\textwidth]{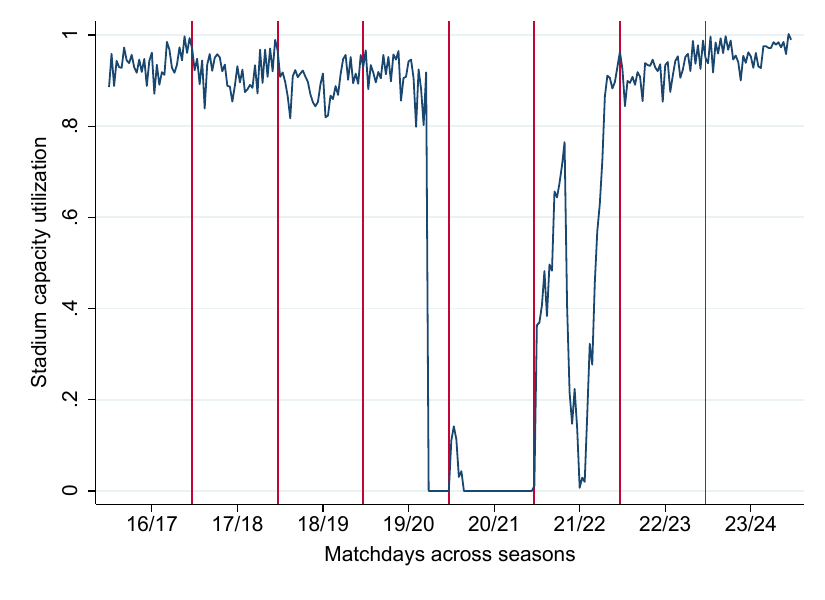}
\subcaption*{Data: \textit{kicker.de}, Bundesliga, first division, eight seasons from 2016/17 to 2023/24. Figure shows the (matchday-)average stadium utilisation rate (spectators/stadium capacity) across 34 matchdays and seasons. Nine matches per matchday. Two ghost games in the season 2019/20 played before matchday 26 are not included.}
\end{figure}

In Germany, the first ghost games in football took place between May and June 2020 and affected the last nine matchdays of season 2019/20. This first ban period has been examined by several studies. We are the first  to focus on a longer time horizon and consider all three seasons subject to spectator restrictions. More specifically, the restrictions (no spectators) applied to the remaining nine matchdays of season 2019/20, to most matches of season 2020/21 and, varying across matchdays and regions and therefore less strict, to season 2021/22, see Figure~\ref{fig:fig1}.

Using data from the first division of the German Bundesliga between 2016/17 and 2023/24 our analysis also includes three seasons before and two seasons after the spectator regulations, all of which experienced an average stadium utilisation rate of more than 90 percent (see Figure \ref{fig:fig1}). Compared to previous research, we are able to examine whether the effects of reduced spectator on-site support are only temporarily or long-lasting. Due to the varying regulations in season 2021/22, we are further able to investigate the impact of partial attendance on HA. Finally, evaluating the post-COVID-19 seasons will show, whether the HA returned to the pre-COVID-19 level after the spectator regulations are lifted - or even a higher one if the effects of returning spectators are self-reinforcing. 

We analyse various match outcomes (goal difference, goals home team, goals away team, win home team, win away team), thereby controlling for multiple fixed effects (home team, away team, matchday, table ranking difference). In doing so, we confirm previous studies regarding the disappearance of the HA during the first spectator ban (last nine matchdays in season 2019/20). For this period, the winning probability of the home team has been reduced by 13 percentage points after the ban came into force. This drop materialised almost entirely through a reduction in home goals, while away goals increased only slightly. Most remarkably, the HA in season 2020/21 (no spectators in most matches) was very close to the pre-COVID-19 season 2018/19. For season 2021/22 (varying spectator regulations), we  detect a U-shaped relationship between HA and stadium utilisation rate, i.e., compared to the other utilisation categories, the HA goes up considerably for matches with medium stadium utilisation. For the two seasons subsequent to the spectator regulations, the HA was at the pre-COVID-19 level.

In further analyses, we control for match fixed effects and also include various covariates (playing in a European competition before or after the match, having a new coach, weekday, weather information). However, our findings remain robust. Moreover, we find considerable heterogeneity in the effect of the first ban period across home and away teams. However, this is unrelated to the stadium capacity, the stadium utilisation rate, the ranking of the teams or the fixed effects of the home and away team. The drop in the HA in the first ban period is only slightly affected when we include several differences between the home and the away team in performance metrics (e.g. shots on goal, corners, fouls). Conversely, we detect that the inverse U-shape relationship between stadium utilisation and HA in season 2021/22 is due to a larger difference in the running distance between the home and away team in matches with a medium utilisation rate.

Our paper contributes to the literature on the HA in football and, more specifically, to the literature testing the effect of spectator absence due to the COVID-19 outbreak. As already indicated above, overall, research has found a substantial drop in the HA for the German Bundesliga's first division. This finding, however, cannot be observed for all major leagues in Europe (\citeauthor{wang-qin-2023}, \citeyear{wang-qin-2023}). While studies looking at several European leagues pooled tend to find a reduction in HA, studies looking at several European leagues separately mostly find a reduction only for the German Bundesliga and few other leagues (see. Appendix Tables~\ref{table:lit1} \& \ref{table:lit2}).

Several studies have examined primary outcomes, i.e. winning percentage, goals, goal difference, points and point difference for the German Bundesliga. It has been found that the spectator ban in season 2019/ 20 was associated with a reduced winning probability of the home team (\citeauthor{dilger-vischer-2022}, \citeyear{dilger-vischer-2022}; \citeauthor{fischer-haucap-2021}, \citeyear{fischer-haucap-2021}; \citeauthor{link-anzer-2022}, \citeyear{link-anzer-2022}; \citeauthor{tilp-thaller-2020}, \citeyear{tilp-thaller-2020}) and an increased winning probability of the away team (\citeauthor{link-anzer-2022}, \citeyear{link-anzer-2022}). \cite{fischer-haucap-2021} further investigated the home winning probability separately for the first, second, and third division and could only detect a reduction for the first division. Although they are only looking at season 2019/20, they already suggest a shrinking effect over the nine affected matchdays. The spectator ban in season 2019/20 was also associated with fewer home goals (\citeauthor{dilger-vischer-2022}, \citeyear{dilger-vischer-2022}) and points (\citeauthor{dilger-vischer-2022}, \citeyear{dilger-vischer-2022}; \citeauthor{tilp-thaller-2020}, \citeyear{tilp-thaller-2020}) as well as a decreased goal difference between home and away goals generally (\citeauthor{link-anzer-2022}, \citeyear{link-anzer-2022}) and within the second half of the match (\citeauthor{santana-et-al-2021}, \citeyear{santana-et-al-2021}). 

There are also studies looking at secondary outcomes related to player and referee behaviour. For the home team reductions in running distance (\citeauthor{dilger-vischer-2022}, \citeyear{dilger-vischer-2022}), shots (\citeauthor{fischer-haucap-2021}, \citeyear{fischer-haucap-2021}), shots on target (\citeauthor{dilger-vischer-2022}, \citeyear{dilger-vischer-2022}) as well as increases in fouls (\citeauthor{fischer-haucap-2021}, \citeyear{fischer-haucap-2021}) and yellow cards (\citeauthor{fischer-haucap-2021}, \citeyear{fischer-haucap-2021}) have been found. In contrast, for the away team, increases in pass accuracy (\citeauthor{dilger-vischer-2022}, \citeyear{dilger-vischer-2022}) and decreases in fouls (\citeauthor{dilger-vischer-2022}, \citeyear{dilger-vischer-2022}) and cards (\citeauthor{dilger-vischer-2022}, \citeyear{dilger-vischer-2022}) have been found. Studies analysing the differences between home and away characteristics are in line with these relations. For more details, we refer to our Appendix Tables~\ref{table:lit1} to \ref{table:lit3}.

The paper proceeds as follows.
Section~\ref{sec:data} describes the dataset and presents descriptive statistics.
Section~\ref{sec:specification} introduces the econometric specification.
Section~\ref{sec:Results} reports and discusses the baseline regression results and presents robustness checks.
Section~\ref{sec:het} analyses whether the effect on the home advantage is heterogeneous and whether there is evidence for potential mechanisms. 
Section~\ref{sec:conclusion} concludes.

\section{Data} \label{sec:data}

Our data consists of all matches of the first division of the German Bundesliga for the eight seasons from 2016/17 to 2023/24 and has been scraped from \textit{kicker.de}, the online-version of a German football magazine. Since this division consists of 18 teams, there are nine matches on each of the 34 matchdays within a season. Hence, each season has a total of 306 matches, i.e., fixtures between a specific home team and a specific away team.\footnote{Consequently, for example, Bayern Munich $-$ Borussia Dortmund and Borussia Dortmund $-$ Bayern Munich are distinct matches.} We drop two ghost matches from season 2019/20 because these were postponed matches belonging to matchdays before the ban came into force, resulting in 2,446 observations at the match-level.\footnote{The match Borussia M{\"o}nchengladbach $-$  1.FC K{\"o}ln on matchday 21 (Feburary 7 $-$ Februar 9,  2020) was postponed to March 11 and was the first ghost match taking place. The match Werder Bremen $-$ Eintracht Frankfurt on matchday 24 (February 28 $-$ March 1, 2020) was postponed to May 3.} Due to relegation and promotion, we observe 28 teams and 652 distinct matches in our dataset across the eight seasons. 

Table~\ref{table:desc} reports descriptive statistics, separately by home and away team. The first two rows indicate a clear HA across the sample period. 45.5\% of all matches ended with a home win, while only 30.2\% ended with an away win (the remaining 24.3\% ended in draws). Correspondingly, home teams scored on average 0.38 goals more than away teams. There are also pronounced differences in performance metrics of the teams. On average, home teams made more shots on goal, delivered more crosses, had more corners, covered a larger running distance, played more passes, completed more dribbles, committed fewer fouls and had less instances of handball, and were offside more often. All of these metrics may influence the number of goals scored and, therefore, the winning probability which we will also examine in Section~\ref{sec:het} where we investigate whether a change in the HA is associated with a change in these metrics.

\begin{table}[ht!]
\centering
\caption{Descriptive statistics by home and away team \label{table:desc}}
\begin{threeparttable}
\fontsize{10}{16}\selectfont
\begin{tabular}{l rl c rl  c }
 \hline

             & \multicolumn{2}{c}{Home team} && \multicolumn{2}{c}{Away team} & $t$-test \\
             \cline{2-3}
             \cline{5-6}
             & Mean & Std. dev.    && Mean & Std. dev. & (equality) \\ [0.5ex]
 \hline
Win (dummy)                       & 0.450   &          && 0.302   &         & $^{***}$ \\
Goals                             & 1.727   & (1.405)  && 1.347   & (1.240) &$^{***}$\\[2ex]
Shots on goal                     & 14.179  & (5.185)  && 11.788  & (4.894) &$^{***}$\\
Crosses                           & 14.186  & (6.523)  && 12.031  & (5.791) & $^{***}$\\
Crosses quota (received/played)   & 0.232    & (0.127) && 0.231    & (0.141)  & \\
Corners                           & 5.217   & (2.900)  && 4.430   & (2.544) &$^{***}$\\
Running distance (km)             & 116.048 & (4.644)  &&  115.729 & (4.834) &  $^{***}$\\
Passes                            &  466.462& (128.8)&& 441.934 &   (119.5)& $^{***}$\\
Dribblings                        &  16.751 & (6.443)  && 15.934  & (6.128) &$^{***}$\\
Ball possesion (share)            & 0.513   & (0.116)  && 0.487   & (0.116) & $^{***}$ \\
Tackle quota (succesful/all)      & 0.506   & (0.052)  && 0.494   & (0.052) & $^{***}$\\
Fouls performed                   & 11.618  & (3.848)  && 12.096  & (3.949) &$^{***}$\\
Offsides                          & 2.109   & (1.684)  && 1.871   & (1.534) &$^{***}$\\
\hline  
\small Observations & \multicolumn{2}{c}{2,446} && \multicolumn{2}{c}{2,446} & \\
\hline
\end{tabular}
\begin{tablenotes}
\footnotesize
\item[] Data: \textit{kicker.de}, Bundesliga, first division, eight seasons from 2016/17 to 2023/24. Two ghost games in the season 2019/20 before matchday 26 are not included. Last column reports significance of $t$-test statistic on equality of groups means. Significance level: $^{*}p<0.05$, $^{**}p<0.01$, $^{***}p<0.001$.
\end{tablenotes}
\end{threeparttable}
\end{table}

\begin{figure}[ht!]
\caption{Match outcomes by season}
\label{fig:outcomes_times}\includegraphics[width=1\textwidth]{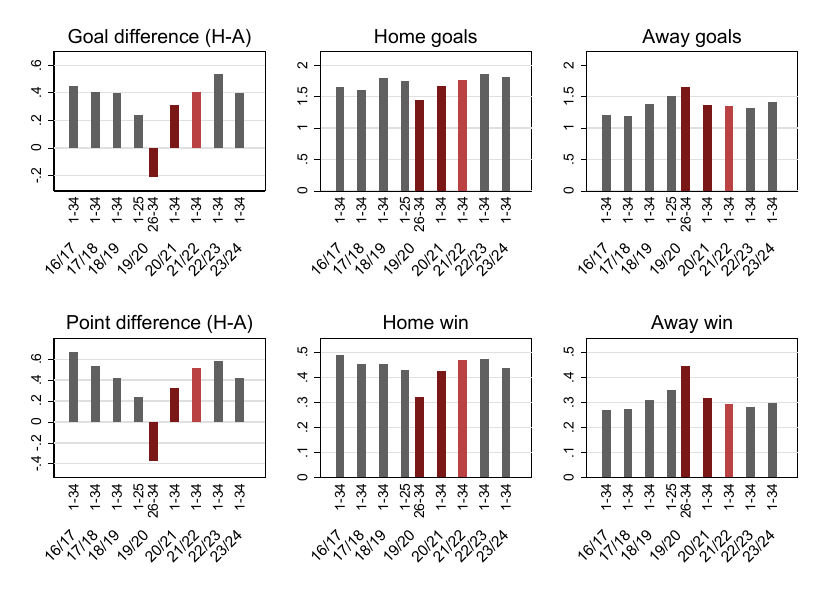}
\subcaption*{Data: \textit{kicker.de}, Bundesliga, first division, eight seasons from 2016/17 to 2023/24. Two ghost games in the season 2019/20 before matchday 26 are not included. There is one bar for each season except for 2019/2020 which is split into matchdays before and after the spectator ban. The three red bars indicate periods with spectator bans (dark red) or attendance restrictions (light red) due to COVID-19. The winning team obtains three points, the loosing team zero points. In case of a draw, both teams obtain one point.}
\end{figure}

Figure~\ref{fig:outcomes_times} displays how the match outcomes have evolved over the eight seasons, thereby, the dark red bars indicate that the spectators have been banned (season 2019/20 matchday 26--34, season 2020/21) and the light red bars indicate that they have been restricted (season 2021/22) due to the COVID-19 outbreak.  
Starting with the goal difference between the home and the away teams (top left), several interesting facts emerge. Before and after the COVID-19 attendance regulations, HA is clearly present. The average goal difference of 0.4 in the last season (2023/24) is almost identical to those of the first three seasons. When the spectator ban came into force (season 2019/20, matchday 26--34) the HA not only fell, but even turned into a disadvantage with an average goal difference of $-0.2$. 
Strikingly, although spectators were still banned during the season 2020/21, the goal difference has been restored to almost pre-COVID-19 levels. While only descriptively, this is suggestive that teams became used to the absence of spectators and possibly adjusted their playing behaviour. \footnote{An alternative interpretation could be that the HA in earlier years has hardly been driven by spectators and that the drop in HA during the matchdays 26-34 of the 2019/2020 season was a general lock-down effect which had an impact on all aspects of life such that the other mechanisms of the HA became less influential.} Relatedly, for season 2021/22 when average stadium\_utilisation was only at 50\%, the goal difference is identical to the seasons before Corona.

The remaining two graphs in the upper panel of Figure~\ref{fig:outcomes_times} decompose the goal difference into goals of the home and goals of the away team. It is evident that the reduction in the goal difference in the no-spectator matches in season 2019/20 is due to less home goals and (to a slightly weaker extent) to more away goals. The lower panel of  Figure~\label{fig:stad_util} shows that the the aforementioned observations on the evolution of goals also hold for the points difference, the probability of a home win and the probability of an away win.\footnote{The winning team obtains three points, the loosing team zero points. In case of a draw, both teams obtain one point.} In particular, in the first period without spectators (matchdays 26-34 of season 2019/20) the probability of an away win exceeds the probability of a home win by 12 percentage points. In all other periods,including season 2020/21 when spectators were banned and season 2021/22 when the number of spectators was restricted, it is significantly more likely that the home team wins. While we think these are interesting results, they are only descriptive. We will examine in the following whether the findings remain after controlling for several potential confounding factors.

\section{Empirical strategy} \label{sec:specification}
We estimate the following equation of outcome $Y_ {is}$ of match $i$ in season $s$ on matchday $j$:
\begin{flalign}
Y_{is} = \beta_0 +  \sum_{k, k\neq 2018/19} \beta_{1,k} S_{k(is)} + \beta_2  S_{2019/20(is)} \times M_{26-34(is)}  \nonumber  \\
+   \gamma_1  S_{2021/22(is)} \times SU_{(0, 0.25]} 
+   \gamma_2  S_{2021/22(is)} \times SU_{(0.25, 0.75]} 
\nonumber \\[1ex]
+   \delta_{MD_j(is)}
+   \alpha_{1,H(i)} + \alpha_{2,A(i)} +  \delta_{Ranking(H-A)_{is}}  + \epsilon_{is} 
\label{eq:baseline}
\end{flalign}
where $Y$ is either (i) the difference in the goals scored by the home team and those scored by the away team, (ii) the goals scored by the home team, (iii) the the goals scored by the away team, (iv) a dummy variable indicating that the home team has won or (v) a dummy variable indicating that the away team has won. $S_k$ denote season dummies whose coefficients measure the difference in the outcome variable in the respective season compared to the baseline season 2018/19. 
$S_{2019/20} \times M_{26-34}$ is equal to one for matchdays 26-34 in season 2019/20, i.e., for those matchdays in season 2019/20 that were affected by the ban for spectators. Hence, $\beta_2$ represents the effect of the ban. To ensure that the estimate of $\beta_2$ is not confounded by the match outcome being generally different in the last part of the season, we include matchday fixed effects $\delta_{MD_j}$. 

Except for 37 matches with a low number of spectators present, spectators were banned throughout season 2020/21. Therefore, we include only one parameter, $\beta_{1,2020}$,  for season 2020/21. \footnote{In a robustness check, we examine whether the match outcome is different for these 37 matches.} During the following season 2021/22, however, stadium utilisation varied markedly (see Figure~\ref{fig:fig1}). For that season we allow the outcome to vary between low (not more than 25\%, indicated by the dummy $SU_{(0, 0.25]}$), medium (between 25 and 75\%, indicated by the dummy $SU_{(0.25, 0.75]}$) and high stadium utilisation rate (above 75\%, the reference category). Thus, \textit{ceteris paribus}, $\gamma_1$ captures the difference in $Y$ between low and high stadium utilisation matches in season 2021/22 while $\beta_{1,2021/22}$ measures the difference between matches with high stadium utilisation in 2021/22 and matches in the reference season 2018/19 (when 83 percent had a high stadium utilisation).

We also incorporate fixed effects for the home team $\alpha_{1,H}$ as well as for the away team $\alpha_{2,A}$. These are not restricted to have just reversed signs, i.e., the relative performance of a certain team can differ when playing at home and away. To control for differences in performance between both teams during the current season, we include fixed effects for the difference in the (before current matchday) table position between the home and away team. $\epsilon$ captures unobserved heterogeneity. 

In further specifications, we also include a vector of control variables comprising information on whether the home respectively away team took part in a European competition between the last matchday and the current one, on whether the home respectively away team will take part in a European competition between the current matchday and the next one, whether the home respectively away team has appointed a new coach since the last matchday, indicators for the weekday on which the match takes place as well as information on relative moisture, precipitation duration and the average temperature during the match.\footnote{Since we are not aware of any systematic data source on the weather conditions during the Bundesliga matches, we measure them by matching the weather conditions of the closest weather station to the respective stadium. In case the closest weather station had missing data, we filled these gaps by using the second or third closest weather station. We, therefore, relied on the every 10 minute data on temperature and precipitation from the \textit{Deutscher Wetterdienst, DWD}. For temperature and relative moisture we use the mean of the nine measurements during the normal time of the game and for precipitation duration and height we use the sum.}

We also run a specification with match fixed effects, i.e., fixed effects for the 652 distinct fixtures $i$, pairing one home and one away team. 202 of these fixtures are only observed once (\textit{singletons}) since (i) five teams have only been in the Bundesliga for one year and there is only a one-year overlap for some teams.  The \textit{singletons} are dropped from the match fixed effects regressions since they do not add to the identification of the coefficients.


\section{Results} \label{sec:Results}

\subsection{Baseline findings}

We start the regression analysis utilising the goal difference between the home and away teams as the outcome variable. The results are presented in Table~\ref{table:goal_diff}. Column (1) reports the coefficients from a specification with seven season dummies (reference category 2018/19), an interaction variable between season 2019/20 and matchday 26-34 (dummy) and a constant. Hence, it reproduces the descriptive result displayed in the top left of Figure~\ref{fig:outcomes_times}.\footnote{For example, the goal difference during the first period of the spectator ban is given by $0.402 - 0.164 - 0.448 = -0.210$.} In the second column, fixed effects for the home as well as for the away team are added, and column (3) shows a specification which additionally incorporates fixed effects for the matchday and for the ranking between the respective opponents.\footnote{In specification (2), the largest home fixed effects are obtained for Bayern Munich (1.83), Borussia Dortmund (1.06), RB Leipzig (0.74), and Bayer Leverkusen (0.32); the largest away fixed effects are obtained for 1. FC N{\"u}rnberg (1.74),  SpVGG Greuther F{\"u}rth (1.70), SV Darmstad 98 (1.11) and VFL Bochum (1.01). Both, the home and the away fixed effects are standardized to sum up to zero.} While the explanatory power increases considerably after the inclusion of the fixed effects, the main pattern stays the same and, therefore, confirms the descriptive findings discussed in Section~\ref{sec:data}. According to specification (3) of Table~\ref{table:goal_diff}, \textit{ceteris paribus} the average goal difference between the home and the away team was 0.35 in the pre-COVID-19 season 2018/19 and was neither statistically different to the two seasons before, nor to the first two thirds of season 2019/20.   

\begin{table}[htbp]
\centering
\def\sym#1{\ifmmode^{#1}\else\(^{#1}\)\fi}
\caption{Goal difference regressions}
\label{table:goal_diff}
    \begin{threeparttable}
    \fontsize{10}{16}\selectfont
    \begin{tabular}{l*{4}{d{3}l}}
    \toprule
        &\multicolumn{2}{c}{(1)}
        &\multicolumn{2}{c}{(2)}
        &\multicolumn{2}{c}{(3)}
        &\multicolumn{2}{c}{(4)}              \\
&\multicolumn{8}{c}{\textbf{Dependent variable: goals home team $-$ goals away team}} \\
\midrule
\multicolumn{9}{l}{\textbf{Season dummies, reference: season 18/19}}\\
 \quad S16/17                             &       0.046         &     (0.162)&       0.192         &     (0.164)&       0.173         &     (0.164)&       0.173         &     (0.164)   \\
\quad S17/18                             &       0.007         &     (0.159)&       0.062         &     (0.155)&       0.047         &     (0.155)&       0.047         &     (0.155)  \\
\quad S19/20                             &      -0.164         &     (0.192)&      -0.037         &     (0.185)&      -0.026         &     (0.185)&      -0.020         &     (0.185)  \\
\quad S20/21                             &      -0.082         &     (0.165)&      -0.004         &     (0.161)&       0.006         &     (0.161)&       0.009         &     (0.161)  \\
\quad S21/22                             &       0.003         &     (0.170)&      -0.007         &     (0.178)&      -0.007         &     (0.179)&      -0.246         &     (0.252)  \\
\quad S22/23                             &       0.137         &     (0.170)&       0.169         &     (0.166)&       0.161         &     (0.165)&       0.162         &     (0.165)   \\
\quad S23/24                             &      -0.007         &     (0.170)&       0.016         &     (0.174)&       0.022         &     (0.172)&       0.024         &     (0.172)   \\
\multicolumn{9}{l}{\textbf{S19/20 $\times$ matchday 26-34} } \\
                                         &      -0.448         &     (0.287)&      -0.443\sym{*}  &     (0.249)&      -0.502\sym{*}  &     (0.268)&      -0.524\sym{*}  &     (0.268)  \\
\multicolumn{9}{l}{\textbf{S21/22 $\times$ stadium utilisation}}\\
\quad \qquad \quad \quad $[0,0.25]$        &                     &            &                     &            &                     &            &      -0.101         &     (0.330)  \\
\quad \qquad \quad \quad  $(0.25,0.75]$   &                     &            &                     &            &                     &            &       0.560\sym{**} &     (0.255)\\
\textbf{Constant}                                 &       0.402\sym{***}&     (0.123)&       0.346\sym{***}&     (0.122)&       0.349\sym{***}&     (0.121)&       0.347\sym{***}&     (0.121)  \\[1ex]
\midrule
\textbf{Fixed Effects} \\
\quad Home team            &&&\multicolumn{2}{c}{\checkmark}&\multicolumn{2}{c}{\checkmark}&\multicolumn{2}{c}{\checkmark}  \\
\quad Away team            &&&\multicolumn{2}{c}{\checkmark}&\multicolumn{2}{c}{\checkmark}&\multicolumn{2}{c}{\checkmark}  \\
\quad Matchday             &&&\multicolumn{2}{c}{}&\multicolumn{2}{c}{\checkmark}&\multicolumn{2}{c}{\checkmark}  \\
\quad Ranking difference        &&&\multicolumn{2}{c}{}&\multicolumn{2}{c}{\checkmark}&\multicolumn{2}{c}{\checkmark}  \\[-1ex]
\quad (home $-$ away team) \\
\midrule
Observations          &\multicolumn{2}{c}{2,446} &\multicolumn{2}{c}{2,446}  &\multicolumn{2}{c}{2,446} &\multicolumn{2}{c}{2,446}\\
$R^2$                  &\multicolumn{2}{c}{0.004} &\multicolumn{2}{c}{0.223}  &\multicolumn{2}{c}{0.249} &\multicolumn{2}{c}{0.252}\\
\bottomrule
\end{tabular}
\begin{tablenotes}
    \item Data: \textit{kicker.de}; Bundesliga, first division, eight seasons from 2016/17 to 2023/24. Two ghost games before matchday 26 in the season 2019/20 are not included. Significance level: $^{*}p<0.10$, $^{**}p<0.05$, $^{***}p<0.01$. Heteroscedasticity-robust standard errors in parentheses.
    \end{tablenotes}
    \end{threeparttable}
    \end{table}

However, the goal difference experienced a considerable reduction by 0.5 during the first spectator ban period, i.e, during the last nine matchdays of season 2019/20. Thus, the disappearance of the HA during the first COVID-19 period was not driven by incidental circumstances during the last nine matchdays, by particular teams being more likely to play at home/away in that period or by ranking constellations making it less likely that the home team wins. In the following season (2020/21), when spectators were still completely banned (except for a few games at the beginning of the season), the goal difference was literally the same as in the pre-COVID-19 season. The combination of (i) a sharp reduction in the HA in the first ban period with (ii) no effect on the HA in the second ban period seems a remarkable result. Possible explanations are that the initial drop was not due to the zero attendance rate or teams became used to ghost games and other factors of the HA became more prominent. 

Specification (4) replaces the season 2021/22 dummy by three binary variables indicating the stadium utilisation rate. Interestingly, the relationship between HA and the stadium utilisation rate follows an inverse U-shape. Compared to the other utilisation categories, the HA goes up considerably for matches with medium stadium utilisation.\footnote{Note that all matches of matchdays $26-34$ in season 2019/20  as well as all matches in season 2020/21 fall into the low utilisation category. Regarding the pre- and post-COVID-19 seasons, there are no matches in the low utilisation category. Including an (additional) dummy for matches in one of the two remaining categories in theses seasons turns out to be insignificant (not reported). Hence, the different effects between high and medium capacity utilisation on match outcomes prevails only in season 2021/22.} This suggests that spectators have a pronounced effect when teams were not used to them. We treat specification (4) as our baseline. In the following, we have a look at further outcome variables.

The first two columns of Table~\ref{table:diff_outcomes} decompose the goal difference effect into home and away goal effects. Apparently, the drop in the HA during the last nine matches of season 2019/20 materialises almost entirely through a reduction in home goals, while the away goals increase only slightly. Correspondingly, the jump in the HA for matches with medium stadium utilisation in season 2021/22 is to two thirds driven by an increase in home goals. Notably, season 2020/21 (with no spectators) was very close to the pre-COVID-19 season 2018/19 in both, the average number of home goals as well as in the average number of away goals.  

The next two columns of Table~\ref{table:diff_outcomes} report the probability of a home team win (column 3) respectively an away team win (column 4). During the first ban period, the former reduces by 13 percentage points, while the latter increases by 10 percentage points. Hence, the probability of a draw does only slightly increase. Again, compared to the pre-COVID-19 season 2018/19, season 2020/21 did neither differ in the home team, nor in the away team winning probability. Finally, the jump in the HA for matches with medium stadium utilisation in season 2021/22  manifests in a 12-13 percentage points larger winning probability of the home team (compared to the other two utilisation categories). All subsequent regression tables contain the same outcome variables as Table~\ref{table:diff_outcomes}.

\begin{table}[htbp]\centering
\renewcommand{\arraystretch}{1.25}
\def\sym#1{\ifmmode^{#1}\else\(^{#1}\)\fi}
\caption{Baseline, different outcome variables }
\label{table:diff_outcomes}
\begin{threeparttable}
\fontsize{10}{12}\selectfont 
\begin{tabular}{l *{4}{d{3}l}}
\toprule
           &\multicolumn{2}{c}{(1)}
           &\multicolumn{2}{c}{(2)}
           &\multicolumn{2}{c}{(3)}
           &\multicolumn{2}{c}{(4)}              \\
\multicolumn{1}{r}{\textbf{Dependent variable:}}
           &\multicolumn{2}{c}{\textbf{Goals}}
           &\multicolumn{2}{c}{\textbf{Goals}}
           &\multicolumn{2}{c}{\textbf{Win}}
           &\multicolumn{2}{c}{\textbf{Win}}         \\[-1ex]
 &\multicolumn{2}{c}{\textbf{home  team}}
           &\multicolumn{2}{c}{\textbf{away team}}
           &\multicolumn{2}{c}{\textbf{home team}}
           &\multicolumn{2}{c}{\textbf{away team}}         \\
\midrule
\multicolumn{9}{l}{\textbf{Season dummies, reference: season 18/19}}\\
\quad S16/17                             &   -0.078         &     (0.122)&      -0.252\sym{**} &     (0.109)&       0.057         &     (0.043)&      -0.078\sym{**} &     (0.039)       \\
\quad S17/18                             &  -0.170         &     (0.112)&      -0.217\sym{**} &     (0.101)&      -0.001         &     (0.041)&      -0.045         &     (0.037)        \\
\quad S19/20                             &   0.041         &     (0.127)&       0.072         &     (0.117)&       0.008         &     (0.045)&       0.022         &     (0.042)        \\
\quad S20/21                             &  -0.016         &     (0.117)&      -0.025         &     (0.103)&      -0.014         &     (0.041)&      -0.006         &     (0.038)        \\
\quad S21/22                             &  -0.250         &     (0.178)&      -0.003         &     (0.159)&      -0.041         &     (0.062)&       0.012         &     (0.060)        \\
\quad S22/23                             &   0.095         &     (0.120)&      -0.068         &     (0.105)&       0.018         &     (0.042)&      -0.031         &     (0.038)        \\
\quad S23/24                             &   0.038         &     (0.126)&       0.013         &     (0.105)&      -0.013         &     (0.044)&      -0.025         &     (0.039)        \\
\multicolumn{9}{l}{\textbf{S19/20 $\times$ matchday 26-34} } \\
                                         &  -0.452\sym{**} &     (0.180)&       0.063         &     (0.176)&      -0.131\sym{**} &     (0.061)&       0.101\sym{*}  &     (0.060)       \\
\multicolumn{9}{l}{\textbf{S21/22 $\times$ stadium utilisation}}\\
\quad \qquad \quad \quad $[0,0.25]$        &   0.307         &     (0.219)&       0.410\sym{*}  &     (0.216)&      -0.015         &     (0.081)&       0.062         &     (0.078)       \\
\quad \qquad \quad \quad  $(0.25,0.75]$   &   0.382\sym{**} &     (0.183)&      -0.179         &     (0.159)&       0.118\sym{*}  &     (0.066)&      -0.076         &     (0.063)       \\
\textbf{Constant}                                 &   1.752\sym{***}&     (0.088)&       1.404\sym{***}&     (0.076)&       0.446\sym{***}&     (0.030)&       0.320\sym{***}&     (0.027)       \\[1ex]
\midrule
\textbf{Fixed Effects} \\
\quad Home team           &\multicolumn{2}{c}{\checkmark}&\multicolumn{2}{c}{\checkmark}&\multicolumn{2}{c}{\checkmark}&\multicolumn{2}{c}{\checkmark}  \\
\quad Away team           &\multicolumn{2}{c}{\checkmark}&\multicolumn{2}{c}{\checkmark}&\multicolumn{2}{c}{\checkmark}&\multicolumn{2}{c}{\checkmark}  \\
\quad Matchday            &\multicolumn{2}{c}{\checkmark}&\multicolumn{2}{c}{\checkmark}&\multicolumn{2}{c}{\checkmark}&\multicolumn{2}{c}{\checkmark}  \\
\quad Ranking difference       &\multicolumn{2}{c}{\checkmark}&\multicolumn{2}{c}{\checkmark}&\multicolumn{2}{c}{\checkmark}&\multicolumn{2}{c}{\checkmark}  \\[-1ex]
\quad (home $-$ away team) \\
\midrule
Observations          &\multicolumn{2}{c}{2,446} &\multicolumn{2}{c}{2,446}  &\multicolumn{2}{c}{2,446} &\multicolumn{2}{c}{2,446}\\
$R^2$                  &\multicolumn{2}{c}{0.203} &\multicolumn{2}{c}{0.179}  &\multicolumn{2}{c}{0.164} &\multicolumn{2}{c}{0.188}\\
\bottomrule
\end{tabular}
\begin{tablenotes}
    \item Data: \textit{kicker.de}; Bundesliga, first division, eight seasons from 2016/17 to 2023/24. Two ghost games before matchday 26 in the season 2019/20 are not included. Significance level: $^{*}p<0.10$, $^{**}p<0.05$, $^{***}p<0.01$. Heteroscedasticity-robust standard errors in parentheses.
    \end{tablenotes}
    \end{threeparttable}
    \end{table}
    
Note that at the beginning of season 2017/18 video assistant referees (VARs) were introduced in order to reduce referee bias and therefore, to enhance fairness. We indeed find that the probability of a home win reduces by 5.8 percentage points between the seasons 2016/17 and 2017/18, but the difference is not statistically significant at conventional levels (standard error of 0.041).\footnote{For a firmer assessment on this issue, earlier seasons would need to be included in order to rule out a pre-trend.} The literature on the effect of the VAR on the HA is also inconclusive. While \cite{dufner-et-al-2023} detect a small drop, a meta analysis by \cite{rogerson-et-al-2024}  could not find a significant reduction of home goals after the introduction of the VAR.

Next, we include additional control variables in the regressions which account for whether a team played in a European competition match close to the current match, for whether a team appointed a new coach since the last match, for the weekday on which the match took place, and for the weather conditions during the match. It may be argued that these variables affect the team performance and hence the match outcome and we want to exclude the possibility that the results discussed above are driven by potential confounders. Appendix Table~\ref{table:desc_add_controls} reports averages of these covariates, separately for seasons without ban as well as  for the three periods with spectator restrictions (season 2019/20 matchday 26--34, season 2020/21, season 2021/22). In fact, there are some pronounced differences between the first ban period and the other seasons. During the first ban period, there were no European matches before or after the respective league match, while in all other periods on average 10 percent of teams had a European match in the previous week and 10 percent of teams had a European match in the following week.\footnote{To mitigate the impact and the spread of the pandemic, the European matches were postponed to August 2020 and played as single-leg matches in Lisbon (Champions League) or in Germany (Europe League).} Moreover, compared to the other subsamples, in the first ban period a considerable larger fraction of matches were schedules between Monday and Thursday and (because the matchdays of the first ban period all took place in May and June) the average temperature was much higher while simultaneously, the precipitation height was significantly larger. 

Table~\ref{table:cov} shows the regression results when the additional covariates are included. Being involved in a European match before or after the matchday has generally no impact on the match outcome. This is consistent with \cite{moffat2020impact} who didn't find any statistically significant impact of Europa League participation during group stage on outcomes of teams from high-performing leagues.
\footnote{We only find a positive and statistically significant effect in the case that the home team had a European match in the following week on the goals of the home team. However, this does not come along with a higher probability of a home team win. Furthermore, (unreported) regression results show the positive effect on the goals of the home team is solely driven by the first four seasons.} Moreover, and perhaps surprisingly, having appointed a new coach since the last match has no effect at all on the winning probabilities. In matches that take place between Monday and Thursday, the goals of the home team reduce on average by 0.25 (compared to matches which take place on a Saturday), but the 4.2 percentage drop in the probability of a home win is not statistically significant. If a match takes places on a Sunday instead of a Saturday, the probability that the away team wins drops by 5 percentage points. While the precipitation duration has no effect on the match outcome, the average temperature does. Compared to the reference temperature category ($5^\circ\mathrm{C} - 15^\circ\mathrm{C}$), the home team scores 0.2 goals less and has a 7.4 percentage point lower winning probability if the temperature is below $5^\circ\mathrm{C}$. Finally, if the temperature is above $15^\circ\mathrm{C}$, the away team scores more goals and has a higher winning probability (compared to the reference temperature category).

\newgeometry{left=10mm, right=10mm, top=10mm, bottom=20mm}
\begin{table}[htbp]\centering
\def\sym#1{\ifmmode^{#1}\else\(^{#1}\)\fi}
\caption{Regressions with covariates\label{table:cov}}
\begin{threeparttable}
\fontsize{10}{16}\selectfont
\begin{tabular}{l *{4}{d{3}l}}
\toprule
           &\multicolumn{2}{c}{(1)}
           &\multicolumn{2}{c}{(2)}
          &\multicolumn{2}{c}{(3)}
           &\multicolumn{2}{c}{(4)}              \\
          &\multicolumn{2}{c}{\textbf{Goals}}
           &\multicolumn{2}{c}{\textbf{Goals}}
           &\multicolumn{2}{c}{\textbf{Win}}
           &\multicolumn{2}{c}{\textbf{Win}}         \\[-1ex]
 &\multicolumn{2}{c}{\textbf{home  team}}
           &\multicolumn{2}{c}{\textbf{away team}}
           &\multicolumn{2}{c}{\textbf{home team}}
           &\multicolumn{2}{c}{\textbf{away team}}         \\
\midrule
\multicolumn{9}{l}{\textbf{Season dummies, reference: season 18/19}}\\
\quad S16/17                             &       -0.079         &     (0.122)&      -0.258\sym{**} &     (0.110)&       0.059         &     (0.044)&      -0.081\sym{**} &     (0.039)        \\
\quad S17/18                             &       -0.126         &     (0.114)&      -0.247\sym{**} &     (0.102)&       0.017         &     (0.042)&      -0.052         &     (0.037)        \\
\quad S19/20                             &        0.021         &     (0.130)&       0.053         &     (0.120)&       0.001         &     (0.046)&       0.012         &     (0.042)        \\
 \quad S20/21                            &        0.014         &     (0.120)&       0.011         &     (0.107)&      -0.005         &     (0.043)&       0.004         &     (0.039)        \\
\quad S21/22                             &       -0.233         &     (0.183)&       0.004         &     (0.166)&      -0.021         &     (0.064)&       0.026         &     (0.062)        \\
\quad S22/23                             &        0.087         &     (0.123)&      -0.096         &     (0.107)&       0.018         &     (0.042)&      -0.040         &     (0.038)        \\
\quad S23/24                             &        0.032         &     (0.127)&      -0.003         &     (0.105)&      -0.013         &     (0.044)&      -0.030         &     (0.039)        \\
\multicolumn{9}{l}{\textbf{S19/20 $\times$ matchday 26-34} } \\
                                         &       -0.374\sym{**} &     (0.189)&      -0.068         &     (0.186)&      -0.116\sym{*}  &     (0.063)&       0.067         &     (0.064)       \\
\multicolumn{9}{l}{\textbf{S21/22 $\times$ stadium utilisation}}\\
\quad \qquad \quad \quad $[0,0.25]$        &        0.356          &     (0.230)&       0.275         &     (0.219)&       0.004         &     (0.085)&       0.016         &     (0.080)      \\
\quad \qquad \quad \quad  $(0.25,0.75]$   &        0.345\sym{*}   &     (0.190)&      -0.169         &     (0.166)&       0.092         &     (0.068)&      -0.078         &     (0.065)      \\
\midrule
\textbf{European match last week}     &\\
\quad Home team                                            &    0.082         &     (0.113)&      -0.066         &     (0.085)&       0.043         &     (0.037)&       0.031         &     (0.031)                  \\
\quad Away team                                            &   -0.012         &     (0.090)&      -0.003         &     (0.098)&      -0.029         &     (0.036)&      -0.037         &     (0.036)                  \\
\textbf{European match next week}     &\\
\quad Home team                                            &    0.191\sym{*}  &     (0.115)&       0.024         &     (0.087)&      -0.015         &     (0.037)&      -0.020         &     (0.031)                  \\
\quad Away team                                            &   0.024         &     (0.094)&      -0.024         &     (0.096)&       0.023         &     (0.036)&       0.022         &     (0.036)\                  \\
\textbf{New coach since last match}   &\\
\quad Home team                                            &   -0.016         &     (0.178)&      -0.305         &     (0.207)&       0.004         &     (0.075)&      -0.004         &     (0.074)               \\
\quad Away team                                           &    0.228         &     (0.243)&       0.275         &     (0.197)&       0.004         &     (0.091)&      -0.003         &     (0.075)              \\
\multicolumn{4}{l}{\textbf{Weekday of the match,
reference: Saturday}}\\
\quad Monday -- Thursday                                   &   -0.252\sym{**} &     (0.119)&       0.174         &     (0.114)&      -0.042         &     (0.046)&       0.061         &     (0.039)              \\
 \quad Friday                                              &    0.144         &     (0.101)&      -0.052         &     (0.084)&       0.046         &     (0.036)&      -0.027         &     (0.031)              \\
 \quad Sunday                                              &    0.014         &     (0.069)&      -0.032         &     (0.066)&       0.002         &     (0.026)&      -0.050\sym{**} &     (0.023)               \\
\textbf{Precipitation duration (min) }    & -0.001          &    (0.001)&      -0.000         &     (0.001)&       0.000         &     (0.001)&       0.000         &     (0.001)                                                                                                                                                  \\
\multicolumn{4}{l}{\textbf{Average temperature during game, reference: $\boldsymbol{[5^\circ\mathrm{C}, 15^\circ\mathrm{C})}$ }}  \\    
 \quad $< 5^\circ\mathrm{C}$                               &   -0.221\sym{***}&     (0.078)&      -0.004         &     (0.070)&      -0.074\sym{**} &     (0.030)&      -0.005         &     (0.027)               \\
 \quad $[15^\circ\mathrm{C}, 25^\circ\mathrm{C})$          &   -0.133         &     (0.089)&       0.265\sym{***}&     (0.079)&      -0.043         &     (0.032)&       0.088\sym{***}&     (0.030)                \\
 \quad $ \ge 25^\circ\mathrm{C}$                          &    0.001         &     (0.199)&       0.234         &     (0.173)&       0.014         &     (0.068)&       0.078         &     (0.064)                \\
\textbf{Constant}                        &        1.796\sym{***} &     (0.097)&       1.360\sym{***}&     (0.084)&       0.461\sym{***}&     (0.034)&       0.310\sym{***}&     (0.030)      \\[1ex]
\midrule
\textbf{Fixed Effects} \\
\quad Home team           &\multicolumn{2}{c}{\checkmark}&\multicolumn{2}{c}{\checkmark}&\multicolumn{2}{c}{\checkmark}&\multicolumn{2}{c}{\checkmark}  \\
\quad Away team           &\multicolumn{2}{c}{\checkmark}&\multicolumn{2}{c}{\checkmark}&\multicolumn{2}{c}{\checkmark}&\multicolumn{2}{c}{\checkmark}  \\
\quad Matchday            &\multicolumn{2}{c}{\checkmark}&\multicolumn{2}{c}{\checkmark}&\multicolumn{2}{c}{\checkmark}&\multicolumn{2}{c}{\checkmark}  \\
\quad Ranking difference       &\multicolumn{2}{c}{\checkmark}&\multicolumn{2}{c}{\checkmark}&\multicolumn{2}{c}{\checkmark}&\multicolumn{2}{c}{\checkmark}  \\[-1ex]
\quad (home $-$ away team) \\
\midrule
Observations          &\multicolumn{2}{c}{2386} &\multicolumn{2}{c}{2386}  &\multicolumn{2}{c}{2386} &\multicolumn{2}{c}{2386}\\
$R^2$                  &\multicolumn{2}{c}{0.203} &\multicolumn{2}{c}{0.179}  &\multicolumn{2}{c}{0.164} &\multicolumn{2}{c}{0.188}\\
\bottomrule
\end{tabular}
\begin{tablenotes}
\footnotesize
    \item Data: \textit{kicker.de}; Bundesliga, first division, eight seasons from 2016/17 to 2023/24. Two ghost games before matchday 26 in the season 2019/20 are not included. Significance level: $^{*}p<0.10$, $^{**}p<0.05$, $^{***}p<0.01$. Heteroscedasticity-robust standard errors in parentheses.
    \end{tablenotes}
    \end{threeparttable}
    \end{table}
    
\restoregeometry

Although some covariates have turned out to affect the match outcome and also vary across seasons, our central findings remain qualitatively unchanged. First, in the first COVID-19 period (season 2019/20, matchdays $26 - 34$), there is a significant drop in the HA regarding both, goals scored by the home team and the winning probability of the home team. Second, season 2020/21 when spectators where also banned does not differ in match outcomes from the pre-COVID-19 seasons. Third, during season 2021/22 with varying spectator regulations we observe a positive effect on the HA for matches with medium stadium utilisation. Fourth, the HA in the post-COVID-19 seasons 2022/23 and 2023/24 is at the same level as in the seasons before the ban came into force.

\begin{table}[htbp]\centering
\def\sym#1{\ifmmode^{#1}\else\(^{#1}\)\fi}
\caption{Inclusion of match fixed effects \label{table:match_FE}}
\begin{threeparttable}
\fontsize{10}{16}\selectfont
\begin{tabular}{l *{4}{d{3}l}}
\toprule
           &\multicolumn{2}{c}{(1)}
           &\multicolumn{2}{c}{(2)}
           &\multicolumn{2}{c}{(3)}
           &\multicolumn{2}{c}{(4)}              \\
\multicolumn{1}{r}{\textbf{Dependent variable:}}
           &\multicolumn{2}{c}{\textbf{Goals}}
           &\multicolumn{2}{c}{\textbf{Goals}}
           &\multicolumn{2}{c}{\textbf{Win}}
           &\multicolumn{2}{c}{\textbf{Win}}         \\[-1ex]
 &\multicolumn{2}{c}{\textbf{home  team}}
           &\multicolumn{2}{c}{\textbf{away team}}
           &\multicolumn{2}{c}{\textbf{home team}}
           &\multicolumn{2}{c}{\textbf{away team}}         \\
\midrule
\multicolumn{9}{l}{\textbf{Season dummies, reference: season 18/19}}\\
\quad S16/17                             &       -0.077         &     (0.124)&      -0.229\sym{**} &     (0.113)&       0.053         &     (0.045)&      -0.072\sym{*}  &     (0.040)       \\
\quad S17/18                             &       -0.172         &     (0.113)&      -0.206\sym{**} &     (0.102)&      -0.005         &     (0.042)&      -0.035         &     (0.037)       \\
\quad S19/20                             &       -0.029         &     (0.133)&       0.096         &     (0.120)&      -0.000         &     (0.046)&       0.024         &     (0.043)       \\
\quad S20/21                             &       -0.019         &     (0.120)&      -0.018         &     (0.109)&      -0.018         &     (0.043)&      -0.004         &     (0.039)       \\
\quad S21/22                             &       -0.111         &     (0.185)&       0.005         &     (0.170)&      -0.095         &     (0.068)&       0.015         &     (0.065)       \\
\quad S22/23                             &        0.084         &     (0.123)&      -0.052         &     (0.109)&       0.011         &     (0.043)&      -0.023         &     (0.039)       \\
\quad S23/24                             &        0.012         &     (0.130)&       0.036         &     (0.109)&      -0.023         &     (0.045)&      -0.018         &     (0.041)       \\
\multicolumn{9}{l}{\textbf{S19/20 $\times$
matchday 26/34 }} \\
                                         &       -0.230         &     (0.203)&       0.004         &     (0.213)&      -0.101         &     (0.076)&       0.107         &     (0.076)      \\
\multicolumn{9}{l}{\textbf{S21/22 $\times$
stadium utilisation}}\\
\quad \qquad \quad \quad $[0,0.25]$        &        0.081         &     (0.239)&       0.524\sym{**} &     (0.243)&       0.021         &     (0.091)&       0.128         &     (0.087)      \\
\quad \qquad \quad \quad  $(0.25,0.75]$    &        0.160         &     (0.207)&      -0.227         &     (0.188)&       0.188\sym{**} &     (0.078)&      -0.100         &     (0.074)      \\
\textbf{Constant }                         &        1.770\sym{***}&     (0.089)&       1.398\sym{***}&     (0.079)&       0.452\sym{***}&     (0.031)&       0.315\sym{***}&     (0.028)      \\[1ex]
\midrule
\textbf{Fixed Effects} \\
\quad Home $\times$ away team           &\multicolumn{2}{c}{\checkmark}&\multicolumn{2}{c}{\checkmark}&\multicolumn{2}{c}{\checkmark}&\multicolumn{2}{c}{\checkmark}  \\
\quad Matchday                           &\multicolumn{2}{c}{\checkmark}&\multicolumn{2}{c}{\checkmark}&\multicolumn{2}{c}{\checkmark}&\multicolumn{2}{c}{\checkmark}  \\
\multicolumn{1}{l}{\quad Ranking difference} &\multicolumn{2}{c}{\checkmark}&\multicolumn{2}{c}{\checkmark}&\multicolumn{2}{c}{\checkmark}&\multicolumn{2}{c}{\checkmark}  \\[-1ex]
\multicolumn{1}{l}{\quad  (home $-$ away team)} \\
\midrule
Observations          &\multicolumn{2}{c}{2,244} &\multicolumn{2}{c}{2,244}  &\multicolumn{2}{c}{2,2447} &\multicolumn{2}{c}{2,244}\\
$R^2$                  &\multicolumn{2}{c}{0.313} &\multicolumn{2}{c}{0.328}  &\multicolumn{2}{c}{0.327} &\multicolumn{2}{c}{0.344}\\
\bottomrule
\end{tabular}
\begin{tablenotes}
\footnotesize
    \item Data: \textit{kicker.de}; Bundesliga, first division, eight seasons from 2016/17 to 2023/24. Two ghost games before matchday 26 in the season 2019/20 are not included. 202 singletons also dropped. Significance level: $^{*}p<0.10$, $^{**}p<0.05$, $^{***}p<0.01$. Heteroscedasticity-robust standard errors in parentheses.
\end{tablenotes}
    \end{threeparttable}
    \end{table}

In the following, we include match fixed effects into our baseline specifications. These are more general than home team and away team fixed effects, as the away team effect for Borussia Dortmund, for example, may be different when playing against FC Schalke 04 compared to playing against VfB Stuttgart. Correspondingly, the $R^2$ increases considerably, e.g., from $0.16$ to $0.33$ in the home win regressions (see Table~\ref{table:match_FE}). The downside, however, is that we lose all the between-fixture variation for identifying the coefficient estimates. Nevertheless, as can be seen from Table~\ref{table:match_FE} we obtain the same qualitative effects as in our baseline specifications. Regarding the drop in the HA in the first ban period (season 2019/20, matchdays 26-34), the inclusion of match fixed effects yields somewhat smaller effect sizes, which have lost their statistical significance. For example, the drop in the probability of a home win is now 10 percentage points (standard error of 0.076) compared to 13.1 percentage  points (standard error of 0.061) in the baseline specification. Regarding the positive HA effect for matches with medium stadium utilisation during season 2021/22, the effect size even increases.\footnote{The medium stadium utilisation category in season 2021/22 does not only exhibit a larger (lower) probability of a win of the home (away) team than the other two stadium utilisation categories in season 2021/22, but also compared to the pre-COVID-19 season 2018/19. Regarding the latter, the medium stadium utilisation category in season 2021/22 has a larger probability of a home team win of 9.3 percentage points (standard error of 0.057) and a lower probability of an away team win of 8.5 percentage points (standard error of 0.051).} 

\subsection{Robustness}

We include several checks to examine the stability of the results. First, we inspect whether our main findings are driven by a particular team and alternately leave out one team from the regressions (i.e., all matches are excluded where the specific team appears either as the home or as the away team). Appendix Figure~\ref{fig:leave_out} shows the corresponding coefficient estimates of the first ban period (season 2019/20, matches $26-34$) in regressions with the dependent variable equal to one if the home team won. It is evident that we obtain a negative estimate in each case which (in absolute values) never falls below nine percentage points.

Second, we drop the last four matchdays from the analysis since one may argue that the determinants of the match outcome differ at the end of the season when the stakes are very high for some teams (for example for those who battle against relegation) while for others the match outcome is more or less meaningless (for example for a team which already got relegated). As we can see from Appendix Table~\ref{table:last4}, the drop in the HA in the first COVID-19 period (season 2019/20, matches $26-34$) becomes even stronger. Without the last four matchdays, the probability that the home team wins falls by 23 percentage points in the first COVID-19 period (13 percentage points in the baseline, see 
Table~\ref{table:diff_outcomes}). 

Third, we drop the 37 matches in season 2020/21 with spectators. On average, the stadium utilisation rate in these matches was 10 percent, with a maximum of 20 percent. Appendix Table~\ref{table:37spec} reports the coefficient estimates of the baseline specifications without these matches. The coefficients on the 2020/21 season dummy remain very small and statistically insignificant.\footnote{Correspondingly, (unreported) regressions using the full sample and including an additional dummy variable indicating these 37 matches, yielded very small and insignificant estimates also for this additional dummy variable.} All other coefficient estimates remain almost identical.\footnote{By construction, without fixed effects they would remain exactly identical.} Hence, we can safely conclude that the surprising finding that the HA in the second COVID-19 period (season 2020/21) was the same as pre-COVID-19 is not due to matches where at least some spectators were allowed.

Fourth, we drop matchdays $18-25$ to rule out that the estimates of the first COVID-19 period (season 2019/20, matchdays $26-34$) are driven by anticipation effects. However, the coefficient estimates hardly change (see Appendix Table~\ref{table:18to25}).

Fifth, in addition to matchdays $18-25$ we also exclude matchdays $1-9$, Hence, the effect of the first COVID-19 period is obtained by comparing the last nine matchdays of the second half with the last nine matchdays of the first half, which may be the better contrast (at the expense of a smaller sample size). However, our main results are not affected by the sample restriction (see Appendix Table~\ref{table:lastnine}). 

Seventh, we estimate an even more general form of Equation~(\ref{eq:baseline}) where we interact the indicator variable for matchday $26-34$ with all seasons (and for season 2021/22 with the three stadium utilisation rates). As documented in Table~\ref{table:postinteract}, again our baseline findings remain fully robust. When comparing with the match outcome in the last nine matches of season 2018/19, we still find a considerable reduction in the HA during the first COVID-19 period. There is no difference between season 2020/21 and the pre-COVID-19 season 2018/19, neither for the first 25 matches nor for the last nine. Regarding the stadium utilisation categories in season 2021/22, a medium utilisation rate increases the HA compared to a high stadium utilisation in the last nine matchdays and compared to matches with a low rate (which occur only in the first 25 matches).\footnote{Regarding a home team win and compared to the probability in a match with high utilisation rate during the last nine matchdays, for example, the probability is 33.9 ($4.8 + 29.1$) percentage points larger in a match with medium utilisation rate during the last nine matches and 30.1  ($6.0 - 5.0 +29.1$) percentage points larger in a match with medium utilisation rate during the first 25 matches. Both differences are statistically significant at the five percent level. Regarding the probability of an away team win and compared to a match with low utilisation rate, for example, the probability is 12.0 ($-5.7 - 6.3$) percentage points lower in a match with medium utilisation rate during the first 25 matchdays, which is statistically significant at the ten percent level.}

\section{Heterogeneities and mechanisms} \label{sec:het}
We have documented above that the HA disappeared in the initial period with ghost games (season 2019/20, matchdays $26-34$). We will examine in this section whether the reduction in the HA was heterogeneous across teams and whether we find evidence for potential mechanisms.

\begin{figure}[ht!]
\caption{Home team heterogeneity of first spectator ban period on home team winning probability}
\label{fig:het_home_win}
\includegraphics[width=0.7 \textwidth]{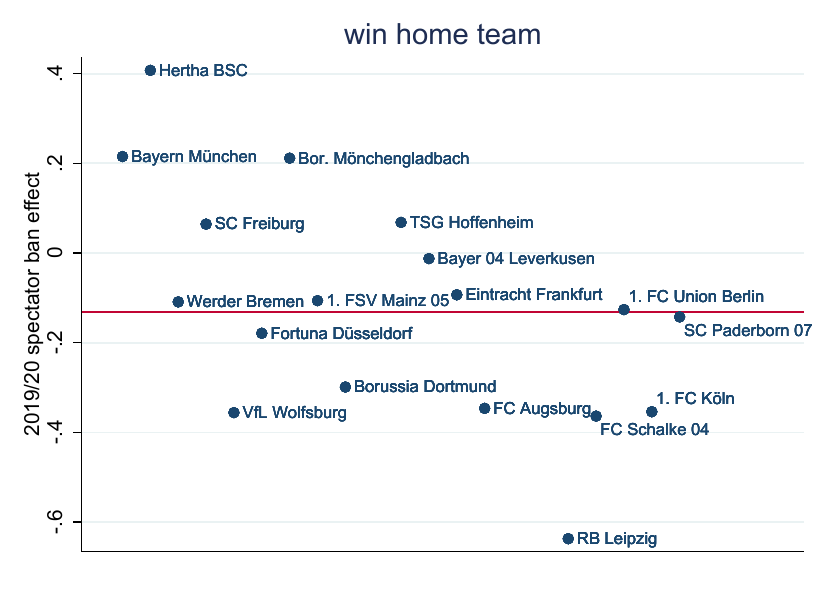}
\subcaption*{\footnotesize Data: \textit{kicker.de}, Bundesliga, first division, eight seasons from 2016/17 to 2023/24. Two ghost games in season 2019/20 before matchday 26 are not included. 
Each point represents for matchdays $26-34$ of the season 2019/20 the (per home team) average difference between the actual value of the home win indicator and the predicted value of the home win indicator, where the predicted values are obtained from estimating Equation~(\ref{eq:baseline}) without  \texttt{S19/20 $\times$ matchday 26-34} for the sample without matchdays $26-34$ of the season 2019/20. The red horizontal line indicates the weighted average effect of $-0.132$. Note that this number marginally deviates from the reported coefficient on \texttt{S19/20 $\times$ matchday 26-34} in Table~\ref{table:diff_outcomes}, specification~(3), since for the latter the fixed effects estimates were identified from the full sample.
}
\end{figure}

Figure~\ref{fig:het_home_win} splits the effect of the first period without spectators on the home win probability by home team, which has been achieved in the following steps. First, Equation~(\ref{eq:baseline}) is estimated for the sample without matchdays $26-34$ of the season 2019/20 (i.., the variable \texttt{S19/20 $\times$ matchday 26-34} is dropped). Second, using these estimates we calculate predicted values on the home win probability of matches taking place during matchdays $26-34$ of season 2019/20. Third, we take the difference between the actual outcome of the home win indicator and the predicted value and calculate then the home team average of these differences. 

There is considerable variation in the 2019/20 ban effect, and  Figure~\ref{fig:het_home_win} shows that for one third of the home teams there is a zero or even a positive impact. A similar spread occurs when plotting the 2019/20 ban effect by away team (see Appendix Figure~\ref{fig:het_away_team_home_win}). However, the average values for a particular team being a home team respectively an away team are not correlated (correlation coefficient of $-0.1$). Further, we do not find any evidence that the effect on the home win probability (or on the goals of the home team) depends on either the ranking of the home team or the away team before the match, on the stadium capacity or the stadium utilisation rate before the ban (see Appendix Figure~\ref{fig:het_combined}).\footnote{There is also no relationship between the 2019/20 ban effect and the fixed effects of the home team or the away team from earlier matches (unreported).}

We have seen in Table~\ref{table:desc} that there are pronounced differences in performance metrics (like shots on goal or corners) between the home and the away team. In the following, differences in these performance measures are included into our baseline specifications to examine whether the change in the effect of the HA is associated with a change in any of these variables. The regression estimates are reported in Table~\ref{table:perf}. As expected, the difference in  the shots on goal between the home and the away team goes along with a higher success of the home team. The difference in the crosses quota is only associated with lower number of goals of the away team, but not with more goals of the home team. Conversely, the difference in the tackles quota is only tied to more goals of the home team. Surprisingly, a larger difference (between the home and the away team) in corners as well as in the passes quota implies a lower success of the home team.\footnote{These two relationships hold not only \textit{ceteris paribus}, but also when all other performance variables are excluded or even in the bivariate case (unreported).} A difference in the running distance of one km between the home and the away team is related to 0.12 more (less) goals of the home (away) team and a five percentage points higher likelihood that a home team win. We find only small or statistically insignificant effects on the match outcome for the differences in performed fouls, in the dribbling quota or in the number of offsides.

Despite several match metrics turning out to be significant in the regressions, most coefficient estimates of the baseline specifications are hardly affected. Even after controlling for all these performance measures, the HA considerably drops in the first COVID period (seasons 2019/20, matchdays  $26-34$). Now, the number of home goals decreases by 0.4 in the first COVID period (compared to 0.45 in the baseline specification)  and the probability of a home team win falls by 10.6 percentage points (compared to 13.1 in the baseline specification).
Again, there is no difference in the match outcomes between the second COVID period (season 2020/21) and the pre-COVID season 2018/19. After the inclusion of the performance variables, however, the probability of a home win or of an away win in the season 2021/22 does not depend any more on the stadium utilisation rate. In fact, the change in the stadium utilisation coefficients compared to the baseline is 
is mainly driven by a difference in the running distance (unreported). The difference in the running distance in the overall sample is 0.3 (see Table~\ref{table:desc}). Yet, in matches with medium stadium utilisation in the season 2021/22 the respective figure is equal to 0.85, while it is close to zero in matches with the other two stadium utilisation categories.

\newgeometry{left=10mm, right=10mm, top=10mm, bottom=20mm}
\begin{table}[htbp]\centering
\def\sym#1{\ifmmode^{#1}\else\(^{#1}\)\fi}
\caption{Inclusion of performance indicators\label{table:perf}}
\begin{threeparttable}
\fontsize{10}{16}\selectfont
\begin{tabular}{l *{4}{d{3}l}}
\toprule
           &\multicolumn{2}{c}{(1)}
           &\multicolumn{2}{c}{(2)}
          &\multicolumn{2}{c}{(3)}
           &\multicolumn{2}{c}{(4)}              \\
          &\multicolumn{2}{c}{\textbf{Goals}}
           &\multicolumn{2}{c}{\textbf{Goals}}
           &\multicolumn{2}{c}{\textbf{Win}}
           &\multicolumn{2}{c}{\textbf{Win}}         \\[-1ex]
 &\multicolumn{2}{c}{\textbf{home  team}}
           &\multicolumn{2}{c}{\textbf{away team}}
           &\multicolumn{2}{c}{\textbf{home team}}
           &\multicolumn{2}{c}{\textbf{away team}}         \\
\midrule
\multicolumn{9}{l}{\textbf{Season dummies, reference: season 18/19}}\\
\quad S16/17                             &    -0.077         &     (0.112)&      -0.224\sym{**} &     (0.100)&       0.051         &     (0.038)&      -0.065\sym{*}  &     (0.035)          \\
\quad S17/18                             &    -0.190\sym{*}  &     (0.104)&      -0.183\sym{**} &     (0.091)&      -0.011         &     (0.036)&      -0.030         &     (0.033)          \\
\quad S19/20                             &     0.085         &     (0.113)&       0.043         &     (0.104)&       0.023         &     (0.039)&       0.012         &     (0.036)          \\
 \quad S20/21                            &     0.027         &     (0.107)&      -0.050         &     (0.093)&      -0.001         &     (0.036)&      -0.015         &     (0.034)          \\
\quad S21/22                             &    -0.140         &     (0.163)&      -0.097         &     (0.145)&       0.006         &     (0.057)&      -0.026         &     (0.057)          \\
\quad S22/23                             &     0.053         &     (0.108)&      -0.018         &     (0.094)&       0.003         &     (0.036)&      -0.014         &     (0.034)          \\
\quad S23/24                             &     0.039         &     (0.115)&       0.039         &     (0.094)&      -0.010         &     (0.038)&      -0.017         &     (0.036) \\       
\textbf{S19/20 $\times$ matchday 26-34} & & & & & & & & \\
                                         &    -0.398\sym{**} &     (0.165)&       0.014         &     (0.154)&      -0.106\sym{*}  &     (0.056)&       0.077         &     (0.050)         \\
\multicolumn{9}{l}{\textbf{S21/22 $\times$ stadium utilisation}} \\
\quad \qquad \quad \quad $[0,0.25]$        &     0.221         &     (0.200)&       0.477\sym{**} &     (0.194)&      -0.045         &     (0.074)&       0.088         &     (0.072)         \\
\quad \qquad \quad \quad $(0.25,0.75]$   &     0.251         &     (0.171)&      -0.034         &     (0.146)&       0.058         &     (0.061)&      -0.021         &     (0.059)         \\
\textbf{Constant}                        &     1.680\sym{***}&     (0.080)&       1.449\sym{***}&     (0.070)&       0.430\sym{***}&     (0.026)&       0.332\sym{***}&     (0.025)         \\[1ex]
\midrule
$\boldsymbol{\Delta}$  \textbf{home team $-$ away team} &\\
\quad Fouls performed                 & 0.010\sym{*}  &     (0.006)&       0.006         &     (0.005)&       0.001         &     (0.002)&       0.004\sym{**} &     (0.002)   \\[1ex]
\quad Shots on goal                   & 0.036\sym{***}&     (0.005)&      -0.026\sym{***}&     (0.004)&       0.010\sym{***}&     (0.002)&      -0.009\sym{***}&     (0.001)   \\[1ex]
\quad Crosses quota \\[-0.5ex]
\quad (received/played)             & 0.016         &     (0.132)&      -0.321\sym{**} &     (0.125)&       0.075         &     (0.045)&      -0.102\sym{**} &     (0.043)   \\[2ex]
\quad Corners                         & -0.052\sym{***}&     (0.008)&       0.025\sym{***}&     (0.007)&      -0.018\sym{***}&     (0.003)&       0.010\sym{***}&     (0.003)   \\[1ex]
\quad Running distance (km)        & 0.125\sym{***}&     (0.008)&      -0.124\sym{***}&     (0.007)&       0.052\sym{***}&     (0.003)&      -0.046\sym{***}&     (0.002)   \\[1ex]
\quad Passes quota \\[-0.5ex]
\quad (received/played)             & -2.309\sym{***}&     (0.348)&       2.842\sym{***}&     (0.326)&      -1.256\sym{***}&     (0.123)&       1.260\sym{***}&     (0.118)   \\[2ex]
\quad Dribblings quota \\[-0.5ex]
\quad (succesful/all)                 & 0.087         &     (0.137)&      -0.144         &     (0.128)&      -0.024         &     (0.049)&       0.019         &     (0.047)   \\[2ex]
\quad Tackles quota \\[-0.5ex]
\quad (successful/all)               & 0.685\sym{**} &     (0.321)&       0.008         &     (0.271)&       0.337\sym{***}&     (0.112)&      -0.086         &     (0.097)   \\[2ex]
\quad Offsides                        & 0.026\sym{**} &     (0.012)&      -0.010         &     (0.010)&       0.007\sym{*}  &     (0.004)&      -0.005         &     (0.004)   \\
\midrule
\textbf{Fixed Effects} \\
\quad Home team           &\multicolumn{2}{c}{\checkmark}&\multicolumn{2}{c}{\checkmark}&\multicolumn{2}{c}{\checkmark}&\multicolumn{2}{c}{\checkmark}  \\
\quad Away team           &\multicolumn{2}{c}{\checkmark}&\multicolumn{2}{c}{\checkmark}&\multicolumn{2}{c}{\checkmark}&\multicolumn{2}{c}{\checkmark}  \\
\quad Matchday            &\multicolumn{2}{c}{\checkmark}&\multicolumn{2}{c}{\checkmark}&\multicolumn{2}{c}{\checkmark}&\multicolumn{2}{c}{\checkmark}  \\
\quad Ranking difference       &\multicolumn{2}{c}{\checkmark}&\multicolumn{2}{c}{\checkmark}&\multicolumn{2}{c}{\checkmark}&\multicolumn{2}{c}{\checkmark}  \\[-1ex]
\quad (home $-$ away team) \\
\midrule
$R^2$                  &\multicolumn{2}{c}{0.338} &\multicolumn{2}{c}{0.337}  &\multicolumn{2}{c}{0.351} &\multicolumn{2}{c}{0.356}\\
\bottomrule
\end{tabular}
\begin{tablenotes}
\footnotesize
    \item Data: \textit{kicker.de}; Bundesliga, first division, eight seasons from 2016/17 to 2023/24. Two ghost games before matchday 26 in the season 2019/20 are not included. Significance level: $^{*}p<0.10$, $^{**}p<0.05$, $^{***}p<0.01$. Heteroscedasticity-robust standard errors in parentheses.
    \end{tablenotes}
    \end{threeparttable}
    \end{table}
    
\restoregeometry

\section{Conclusion} \label{sec:conclusion}

The existence of a HA in sports, the phenomenon whereby the home team performs better than the visiting team, is a well-established fact. In the aftermath of the COVID-19 outbreak, spectators were banned from football matches, which we leverage as a natural experiment to examine the impact of spectators on HA. Using data from the German Bundesliga for the eight seasons between 2016/17 and 2023/24, we are the first to take a longer perspective and cover not only the first but all three seasons subject to spectator regulations. We analyse various match outcomes (goals home team, goals away team, win home team, win away team) and control for fixed effects of the home team, the away team, the difference in the table ranking, the matchday, and (in additional analysis) match fixed effects.

We confirm previous studies regarding the disappearance of HA in the last nine matches of season2019/20. The drop in HA materialises almost entirely through a reduction in home goals. While we find considerable heterogeneity in the effect of the first ban period across home and away teams, it is unrelated to the stadium capacity, the stadium utilisation rate, the ranking of the teams or the fixed effects of the home and the away team. The drop in the HA in the first ban period is only slightly affected when we include several differences between the home and the away team in performance metrics. 

The HA in season 2020/21, with no spectators in most matches, was very close to the one of the pre-COVID-19 season 2018/19, indicating that teams became accustomed to the absence of spectators. In addition, this makes it very unlikely that the drop in HA during the first ban period was due to a reduction in referee bias (towards the home team), which is often brought forward as a mechanism for the impact of spectators. However, if that were the case, we would expect the reduction in referee bias to persist also in the second ban period.

For season 2021/22, with varying attendance regulations, we detect a U-shaped relationship between HA and the stadium utilisation rate, where HA increases considerably for matches with medium stadium utilisation due to a larger difference in running distance between the home and away teams. This suggests that the impact of spectators is particularly pronounced when players have been without spectators for some time. 

In future research, we will examine whether the observed time pattern of the impact of ghost games can also be observed in the second and third German football divisions.  
It may also be worthwhile to examine performance metrics at a more granular level, specifically at the player level, to distinguish between those who played during the first ban period and those who played only in the second. This may allow for further conclusions regarding teams getting accustomed to the absence of spectators.   

\clearpage

\begin{onehalfspacing}
\bibliography{football.bib}
\end{onehalfspacing}

\clearpage

\appendix

\clearpage

\section{Appendix}

\renewcommand*{\thefigure}{\thesection\arabic{figure}}
\renewcommand*{\thetable}{\thesection\arabic{table}}
\setcounter{figure}{0}
\setcounter{table}{0}

\newgeometry{left=2.5cm, right=2.5cm, top=15mm, bottom=20mm}
\begin{table}[ht!]
\fontsize{10}{16}\selectfont
\caption{HA and COVID-19 literature -- Europe (incl. Bundesliga\label{table:lit1}), pooled}
\begin{center}
\begin{tabular}{|m{0.2\textwidth}|m{0.12\textwidth}|m{0.2\textwidth}|m{0.40\textwidth}|}
\hline
\textbf{Study} & \textbf{COVID-19 season} & \textbf{$\Delta$HA} & \textbf{Findings}\\
\hline
\hline
\citeauthor{bhagwandeen-et-al-2024} (\citeyear{bhagwandeen-et-al-2024}) & 2019/20 & $HA_{Euro}$ reduced & points (H)$\downarrow$, yellow cards (H)$\uparrow$\\
\hline
\citeauthor{bilalic-et-al-2021} (\citeyear{bilalic-et-al-2021}) & 2019/20 & $HA_{Euro}$ reduced & points (H-A)$\downarrow$, goals (H-A)$\downarrow$, corners (H-A)$\downarrow$, shots (H-A)$\downarrow$, shots on target (H-A)$\downarrow$, fouls (H-A)$\uparrow$, yellow cards (H-A)$\uparrow$\\
\hline
\citeauthor{cross-uhrig-2023} (\citeyear{cross-uhrig-2023}) & 2019/20 & $HA_{Euro}$ reduced & goals (H-A)$\downarrow$  \\
\hline
\citeauthor{cueva-2020} (\citeyear{cueva-2020}) & 2019/20 & $HA_{Euro}$ reduced & win. prob. (H)$\downarrow$, fouls (H)$\uparrow$, yellow cards (A)$\downarrow$ \\
\hline
\citeauthor{ferraresi-gucciardi-2021} (\citeyear{ferraresi-gucciardi-2021}) & 2019/20 & $HA_{Euro}$ reduced & penalty kick quota (H)$\downarrow$, penalty kick quote (A)$\uparrow$ \\
\hline
\citeauthor{ferraresi-gucciardi-2023} (\citeyear{ferraresi-gucciardi-2023}) & 2019/20 & $HA_{Euro}$ reduced & points (H)$\downarrow$, shots (H)$\downarrow$, shots on target (H)$\downarrow$, corners (H)$\downarrow$, clearances (H)$\uparrow$, interceptions (H)$\downarrow$, fouls (H)$\uparrow$, yellow cards (H)$\uparrow$ \\
\hline
\citeauthor{leitner-richlan-2021} (\citeyear{leitner-richlan-2021}) & 2019/20 & $HA_{Euro}$ reduced & win. prob. (H)$\downarrow$, yellow cards (H)$\uparrow$ \\
\hline
\citeauthor{mccarrick-et-al-2021} (\citeyear{mccarrick-et-al-2021}) & 2019/20 & $HA_{Euro}$ reduced & points (H-A)$\downarrow$, goals (H-A)$\downarrow$, corners (H-A)$\downarrow$, shots (H-A)$\downarrow$, shot on target (H-A)$\downarrow$, fouls (H-A)$\uparrow$, yellow cards (H-A)$\uparrow$\\
\hline
\citeauthor{scoppa-2021} (\citeyear{scoppa-2021}) & 2019/20 & $HA_{Euro}$ reduced & points (H)$\downarrow$, goals (H)$\downarrow$, shots (H)$\downarrow$, shots (H-A)$\downarrow$, shots on target (H)$\downarrow$, shots on target (H-A)$\downarrow$, corners (H)$\downarrow$, corners (H-A)$\downarrow$, fouls (H)$\uparrow$, fouls (A)$\uparrow$, fouls (H-A)$\uparrow$, yellow cards (H)$\uparrow$, yellow cards (A)$\downarrow$, yellow cards (H-A)$\uparrow$, red cards (H)$\uparrow$, red cards (H-A)$\uparrow$  \\
\hline
\citeauthor{sors-et-al-2021} (\citeyear{sors-et-al-2021}) & 2019/20 & $HA_{Euro}$ reduced & win prob. (H)$\downarrow$, win prob. (A)$\uparrow$ \\
\hline
\citeauthor{wunderlich-et-al-2021} (\citeyear{wunderlich-et-al-2021}) & 2019/20 & $HA_{Euro}$ mixed &  shots (H)$\downarrow$, shots on target (H)$\downarrow$, fouls (A)$\downarrow$, yellow cards (A)$\downarrow$, red cards (A)$\downarrow$, but no effect for goals, points\\
\hline
\citeauthor{zheng-et-al-2023} (\citeyear{zheng-et-al-2023}) & 2020/21 & $HA_{Euro}$ mixed & penalty kick miss (H)$\downarrow$, penalty kick quota (A)$\downarrow$ \\
\hline
\end{tabular} \end{center}
    \begin{tablenotes}
        \item[] This table includes studies that have used data from several European countries (incl. the Bundesliga) and depicted pooled results for the relationship between the COVID-19 induced spectator ban and the HA development. The column \textit{$\Delta$HA} summarises the studies findings. The column \textit{Findings} refers to the analysed variables for which the authors could detect a relationship with the COVID-19 induced spectator ban. Arrows signal the direction of the relationship. Parentheses () indicate whether the relationship has been identified with respect to the home team (H), the away team (A), or the difference between the two teams (H-A).
    \end{tablenotes}
\end{table}

\restoregeometry

\newpage

\begin{table}[ht!]
\fontsize{10}{16}\selectfont
\caption{HA and COVID-19 literature -- Europe (incl. Bundesliga (BL)), by league\label{table:lit2}}

\begin{center}
\begin{tabular}{|m{0.2\textwidth}|m{0.12\textwidth}|m{0.2\textwidth}|m{0.40\textwidth}|}
\hline
\textbf{Study} & \textbf{COVID-19 season} & \textbf{$\Delta$HA} & \textbf{Findings w. r. t. BL}\\
\hline
\hline
\citeauthor{almeida-leite-2021} (\citeyear{almeida-leite-2021}) & 2019/20 & $HA_{Euro}$ mixed; $HA_{BL}$ reduced & points (H)$\downarrow$, total shots(H)$\downarrow$, shots on target(H)$\downarrow$, ball possession(H)$\downarrow$\\
\hline
\citeauthor{benz-lopez-2023} (\citeyear{benz-lopez-2023}) & 2019/20 & $HA_{Euro}$ mixed; $HA_{BL}$ reduced & goals (H;A)$\downarrow$ \\
\hline
\citeauthor{bryson-et-al-2021} (\citeyear{bryson-et-al-2021}) & 2019/20 & $HA_{Euro}$ reduced; $HA_{BL}$ reduced & yellow cards (H)$\uparrow$, yellow cards (A)$\downarrow$, yellow cards (H-A)$\uparrow$  \\
\hline
\citeauthor{correiaoliveira-andradesouza-2022} (\citeyear{correiaoliveira-andradesouza-2022}) & 2019/20 & $HA_{Euro}$ mixed; $HA_{BL}$ reduced & points$\downarrow$ for BL1, not BL2 \\
\hline
\citeauthor{hill-vanyerpen-2021} (\citeyear{hill-vanyerpen-2021}) & 2019/20 & $HA_{Euro}$ reduced; $HA_{BL}$ reduced & points (H)$\downarrow$, goals (A)$\uparrow$, yellow cards (H)$\uparrow$  \\
\hline
\citeauthor{jimenezsanchez-lavin-2021} (\citeyear{jimenezsanchez-lavin-2021}) & 2019/20 & $HA_{Euro}$ mixed; $HA_{BL}$ reduced & win prob. (H)$\downarrow$, points (H)$\downarrow$, points (A)$\uparrow$ \\
\hline
\citeauthor{krawczyk-strawinski-2022} (\citeyear{krawczyk-strawinski-2022}) & 2019/20 & only $HA_{BL}$ reduced & win. prob. (H)$\downarrow$, goals (H-A)$\downarrow$\\
\hline
\citeauthor{ramchandani-millar-2021} (\citeyear{ramchandani-millar-2021}) & 2019/20 & $HA_{Euro}$ mixed; $HA_{BL}$ reduced & win prob. (H)$\downarrow$ \\
\hline
\citeauthor{sedeaud-2021} (\citeyear{sedeaud-2021}) & 2019/20 & $HA_{Euro}$ mixed; $HA_{BL}$ reduced & win prob. (H)$\downarrow$, win prob. (A)$\uparrow$\\
\hline
\end{tabular}\end{center}
    \begin{tablenotes}
        \item[] This table includes studies that have used data from several European countries to analyse the relationship between the COVID-19 induced spectator ban and the HA development,  but depicted (at least some) results separately for the German Bundesliga. The column \textit{$\Delta$HA} summarises the studies findings, generally for the analysed leagues and specifically for the Bundesliga. The column \textit{Findings} refers to the analysed variables for which the authors could detect a relationship with the COVID-19 induced spectator ban for the Bundesliga. Arrows signal the direction of the relationship. Parentheses () indicate whether the relationship has been identified with respect to the home team (H), the away team (A), or the difference between the two teams (H-A).
    \end{tablenotes}
\end{table}

\newpage

\begin{table}[ht!]
\fontsize{10}{16}\selectfont
\caption{HA and COVID-19 literature -- Bundesliga (BL) only\label{table:lit3}}
\begin{center}
\begin{tabular}{|m{0.2\textwidth}|m{0.12\textwidth}|m{0.2\textwidth}|m{0.40\textwidth}|}
\hline
\textbf{Study} & \textbf{COVID-19 season} & \textbf{$\Delta$HA} & \textbf{Findings}\\
\hline
\hline
\citeauthor{dilger-vischer-2022} (\citeyear{dilger-vischer-2022}) & 2019/20 & HA reduced & win prob. (H)$\downarrow$, points (H)$\downarrow$, goals (H)$\downarrow$, running distance (H)$\downarrow$, pass accuracy (A)$\uparrow$, shots on target (H)$\downarrow$, cards (A)$\downarrow$, fouls (A)$\downarrow$, overtime $\downarrow$\\
\hline
\citeauthor{dufner-et-al-2023} (\citeyear{dufner-et-al-2023}) & 2019/20; 2020/21 & HA reduced & points (H-A)$\downarrow$, fouls (H-A)$\uparrow$, penalty kicks (H-A)$\downarrow$; but small effect sizes (d) and only compare seasons before VAR vs. seasons after VAR \\
\hline
\citeauthor{endrich-gesche-2020} (\citeyear{endrich-gesche-2020}) & 2019/20 & HA reduced & fouls (H-A)$\uparrow$ , yellow cards (H-A)$\uparrow$\\
\hline
\citeauthor{fischer-haucap-2021} (\citeyear{fischer-haucap-2021}) & 2019/20 & HA reduced & win prob. (H)$\downarrow$ for BL1, not BL2 or BL3, points (H-A)$\downarrow$, fouls (H)$\uparrow$, yellow cards (H)$\uparrow$,  shots (H)$\downarrow$ \\
\hline
\citeauthor{link-anzer-2022} (\citeyear{link-anzer-2022}) & 2019/20 & HA reduced & win prob. (H)$\downarrow$, win prob. (A)$\uparrow$,  goals (H-A)$\downarrow$ \\
\hline
\citeauthor{santana-et-al-2021} (\citeyear{santana-et-al-2021}) & 2019/20 & HA reduced & goals in 2nd half (H-A)$\downarrow$, sprints (H-A)$\downarrow$, fouls (H-A)$\downarrow$\\
\hline
\citeauthor{tilp-thaller-2020} (\citeyear{tilp-thaller-2020}) & 2019/20 & HA reduced & win prob. (H)$\downarrow$, points (H) $\downarrow$, fouls (H-A)$\uparrow$, cards (H-A)$\uparrow$; home disadvantage\\
\hline
\end{tabular}\end{center}
    \begin{tablenotes}
        \item[] This table includes studies that have used data from the German Bundesliga analyse the relationship between the COVID-19 induced spectator ban and the HA development. The column \textit{$\Delta$HA} summarises the studies findings. The column \textit{Findings} refers to the analysed variables for which the authors could detect a relationship with the COVID-19 induced spectator ban. Arrows signal the direction of the relationship. Parentheses () indicate whether the relationship has been identified with respect to the home team (H), the away team (A), or the difference between the two teams (H-A).
    \end{tablenotes}
\end{table}

\newpage

\begin{table}[ht!]
\centering
\caption{Control variables, means across subsamples \label{table:desc_add_controls}}
\begin{threeparttable}
\fontsize{10}{16}\selectfont
\begin{tabular}{l c  c  c  c}
 \hline
&\multicolumn{1}{c}{\textbf{No-ban}}  
&\multicolumn{1}{c}{\textbf{S19/20,}} 
&\multicolumn{1}{c}{\textbf{S20/21}}
&\multicolumn{1}{c}{\textbf{S21/22}} \\
&
&
\multicolumn{1}{c}{\textbf{MD 26-34}}
&
& \\
\hline
\textbf{European match last week}     &\\
\quad Home team                    &  0.114& 0.000 &  0.101   & 0.108                 \\
\quad Away team                    &  0.111& 0.000 &  0.092   & 0.111                 \\
\textbf{European match next week}     &\\
\quad Home team                    &  0.115& 0.000 &  0.101   & 0.101                 \\
\quad Away team                    &  0.110& 0.000 &  0.092   & 0.118                 \\
\textbf{New coach since last match}   &\\
\quad Home team                    &  0.017& 0.012 &  0.020   & 0.007              \\
\quad Away team                   &  0.011& 0.012 &  0.026   & 0.02              \\
\hline
\textbf{Weekday of the match}         &\\
\quad Monday -- Thursday           &  0.065& 0.247 &  0.114   & 0.042             \\
 \quad Friday                      &  0.098& 0.049 &  0.082   & 0.095             \\
\quad Saturday                     &  0.618& 0.556 &  0.598   & 0.621             \\
\quad Sunday                       &  0.220& 0.148 &  0.206   & 0.242              \\
\hline
\textbf{Precipitation duration during game (min)}    &  5.708& 6.173 &  4.673   & 5.186         \\
\textbf{Precipitation height (mm)}          &  0.138& 0.218 &  0.068   & 0.108             \\
\textbf{Average temperature during game ($^\circ$C)}       &  10.914& 18.401 &  7.628   & 10.147      \\
\hline
Observations & 1,753 & 81 & 306 & 306 \\
\hline
\end{tabular}
\begin{tablenotes}
\footnotesize
\item[] Data: \textit{kicker.de}, Bundesliga, eight seasons from 2016/17 to 2023/24. Two ghost games in the season 2019/20 before matchday 26 are not included. No-ban comprises the seasons 2016/17, 2017/18, 2018/19, 2022/23, 2023/34 as well as the first 25 matchdays of the season 2019/20.
\end{tablenotes}
\end{threeparttable}
\end{table}

\begin{table}[htbp]\centering
\def\sym#1{\ifmmode^{#1}\else\(^{#1}\)\fi}
\caption{Without the last four matchdays of a season\label{table:last4}}
\begin{threeparttable}
\fontsize{10}{16}\selectfont
\begin{tabular}{l *{4}{d{3}l}}
\toprule
           &\multicolumn{2}{c}{(1)}
           &\multicolumn{2}{c}{(2)}
          &\multicolumn{2}{c}{(3)}
           &\multicolumn{2}{c}{(4)}              \\
          &\multicolumn{2}{c}{\textbf{Goals}}
           &\multicolumn{2}{c}{\textbf{Goals}}
           &\multicolumn{2}{c}{\textbf{Win}}
           &\multicolumn{2}{c}{\textbf{Win}}         \\[-1ex]
 &\multicolumn{2}{c}{\textbf{home  team}}
           &\multicolumn{2}{c}{\textbf{away team}}
           &\multicolumn{2}{c}{\textbf{home team}}
           &\multicolumn{2}{c}{\textbf{away team}}         \\
\midrule
\multicolumn{9}{l}{\textbf{Season dummies, reference: season 18/19}}\\
\quad S16/17                             &     -0.039         &     (0.126)&      -0.324\sym{***}&     (0.110)&       0.075         &     (0.046)&      -0.090\sym{**} &     (0.041)    \\     
\quad S17/18                             &     -0.181         &     (0.115)&      -0.245\sym{**} &     (0.104)&      -0.009         &     (0.044)&      -0.060         &     (0.039)    \\     
\quad S19/20                             &      0.065         &     (0.129)&       0.084         &     (0.120)&       0.006         &     (0.046)&       0.019         &     (0.043)    \\     
 \quad S20/21                            &     -0.021         &     (0.121)&      -0.012         &     (0.108)&      -0.026         &     (0.044)&      -0.005         &     (0.040)    \\     
\quad S21/22                             &     -0.297         &     (0.219)&      -0.124         &     (0.180)&      -0.006         &     (0.075)&      -0.007         &     (0.070)    \\     
\quad S22/23                             &      0.096         &     (0.125)&      -0.037         &     (0.111)&       0.006         &     (0.044)&      -0.028         &     (0.040)    \\     
\quad S23/24                             &      0.050         &     (0.131)&       0.000         &     (0.110)&      -0.012         &     (0.047)&      -0.022         &     (0.042)    \\     
\multicolumn{9}{l}{\textbf{S19/20 $\times$ matchday 26-34} } \\
                                         &     -0.500\sym{**} &     (0.217)&       0.225         &     (0.221)&      -0.228\sym{***}&     (0.069)&       0.108         &     (0.074)   \\  
\multicolumn{9}{l}{\textbf{S21/22 $\times$ stadium utilisation}}\\ 
\quad \qquad \quad \quad $[0,0.25]$        &      0.358         &     (0.254)&       0.551\sym{**} &     (0.229)&      -0.054         &     (0.091)&       0.087         &     (0.085)   \\      
\quad \qquad \quad \quad  $(0.25,0.75]$   &      0.376\sym{*}  &     (0.224)&      -0.050         &     (0.176)&       0.064         &     (0.077)&      -0.051         &     (0.072)   \\      
\textbf{Constant}                        &      1.712\sym{***}&     (0.089)&       1.385\sym{***}&     (0.080)&       0.446\sym{***}&     (0.032)&       0.326\sym{***}&     (0.028)   \\[1ex] 
\midrule
\textbf{Fixed Effects} \\
\quad Home team           &\multicolumn{2}{c}{\checkmark}&\multicolumn{2}{c}{\checkmark}&\multicolumn{2}{c}{\checkmark}&\multicolumn{2}{c}{\checkmark}  \\
\quad Away team           &\multicolumn{2}{c}{\checkmark}&\multicolumn{2}{c}{\checkmark}&\multicolumn{2}{c}{\checkmark}&\multicolumn{2}{c}{\checkmark}  \\
\quad Matchday            &\multicolumn{2}{c}{\checkmark}&\multicolumn{2}{c}{\checkmark}&\multicolumn{2}{c}{\checkmark}&\multicolumn{2}{c}{\checkmark}  \\
\quad Ranking difference       &\multicolumn{2}{c}{\checkmark}&\multicolumn{2}{c}{\checkmark}&\multicolumn{2}{c}{\checkmark}&\multicolumn{2}{c}{\checkmark}  \\[-1ex]
\quad (home $-$ away team) \\
\midrule
$R^2$                  &\multicolumn{2}{c}{0.209} &\multicolumn{2}{c}{0.194}  &\multicolumn{2}{c}{0.170} &\multicolumn{2}{c}{0.201}\\
\bottomrule
\end{tabular}
\begin{tablenotes}
\footnotesize
    \item Data: \textit{kicker.de}; Bundesliga,  first division,eight seasons from 2016/17 to 2023/24. Two ghost games before matchday 26 in the season 2019/20 are not included. Significance level: $^{*}p<0.10$, $^{**}p<0.05$, $^{***}p<0.01$. Heteroscedasticity-robust standard errors in parentheses.
    \end{tablenotes}
    \end{threeparttable}
    \end{table}

\begin{table}[htbp]\centering
\def\sym#1{\ifmmode^{#1}\else\(^{#1}\)\fi}
\caption{Without 37 matches in the season 2020/21 with spectators\label{table:37spec}}
\begin{threeparttable}
\fontsize{10}{16}\selectfont
\begin{tabular}{l *{4}{d{3}l}}
\toprule
           &\multicolumn{2}{c}{(1)}
           &\multicolumn{2}{c}{(2)}
          &\multicolumn{2}{c}{(3)}
           &\multicolumn{2}{c}{(4)}              \\
          &\multicolumn{2}{c}{\textbf{Goals}}
           &\multicolumn{2}{c}{\textbf{Goals}}
           &\multicolumn{2}{c}{\textbf{Win}}
           &\multicolumn{2}{c}{\textbf{Win}}         \\[-1ex]
 &\multicolumn{2}{c}{\textbf{home  team}}
           &\multicolumn{2}{c}{\textbf{away team}}
           &\multicolumn{2}{c}{\textbf{home team}}
           &\multicolumn{2}{c}{\textbf{away team}}         \\
\midrule
\multicolumn{9}{l}{\textbf{Season dummies, reference: season 18/19}}\\
\quad S16/17                             & -0.078         &     (0.122)&      -0.256\sym{**} &     (0.109)&       0.059         &     (0.043)&      -0.080\sym{**} &     (0.039)    \\       
\quad S17/18                             & -0.171         &     (0.112)&      -0.217\sym{**} &     (0.100)&      -0.000         &     (0.041)&      -0.045         &     (0.037)     \\      
\quad S19/20                             &  0.047         &     (0.128)&       0.062         &     (0.118)&       0.006         &     (0.045)&       0.018         &     (0.042)     \\      
\quad S20/21           & -0.014         &     (0.122)&      -0.017         &     (0.105)&      -0.007         &     (0.043)&      -0.007         &     (0.039)     \\      
\quad S21/22                             & -0.244         &     (0.178)&      -0.002         &     (0.159)&      -0.037         &     (0.062)&       0.009         &     (0.060)     \\      
\quad S22/23                             &  0.096         &     (0.120)&      -0.068         &     (0.105)&       0.018         &     (0.042)&      -0.032         &     (0.038)     \\      
\quad S23/24                             &  0.039         &     (0.126)&       0.011         &     (0.105)&      -0.013         &     (0.044)&      -0.026         &     (0.039)     \\      
\multicolumn{9}{l}{\textbf{S19/20 $\times$ matchday 26-34} } \\
                                         & -0.449\sym{**} &     (0.180)&       0.071         &     (0.177)&      -0.126\sym{**} &     (0.061)&       0.102\sym{*}  &     (0.060)    \\       
\multicolumn{9}{l}{\textbf{S21/22 $\times$ stadium utilisation}}\\                                          
\quad \qquad \quad \quad $[0,0.25]$        &  0.302         &     (0.220)&       0.410\sym{*}  &     (0.216)&      -0.017         &     (0.081)&       0.062         &     (0.078)    \\       
\quad \qquad \quad \quad  $(0.25,0.75]$   &  0.375\sym{**} &     (0.184)&      -0.188         &     (0.159)&       0.115\sym{*}  &     (0.066)&      -0.076         &     (0.063)    \\       
\textbf{Constant}                        &  1.751\sym{***}&     (0.088)&       1.406\sym{***}&     (0.076)&       0.445\sym{***}&     (0.030)&       0.322\sym{***}&     (0.027)    \\[1ex]  
\midrule
\textbf{Fixed Effects} \\
\quad Home team           &\multicolumn{2}{c}{\checkmark}&\multicolumn{2}{c}{\checkmark}&\multicolumn{2}{c}{\checkmark}&\multicolumn{2}{c}{\checkmark}  \\
\quad Away team           &\multicolumn{2}{c}{\checkmark}&\multicolumn{2}{c}{\checkmark}&\multicolumn{2}{c}{\checkmark}&\multicolumn{2}{c}{\checkmark}  \\
\quad Matchday            &\multicolumn{2}{c}{\checkmark}&\multicolumn{2}{c}{\checkmark}&\multicolumn{2}{c}{\checkmark}&\multicolumn{2}{c}{\checkmark}  \\
\quad Ranking difference       &\multicolumn{2}{c}{\checkmark}&\multicolumn{2}{c}{\checkmark}&\multicolumn{2}{c}{\checkmark}&\multicolumn{2}{c}{\checkmark}  \\[-1ex]
\quad (home $-$ away team) \\
\midrule
Observations          &\multicolumn{2}{c}{2,409} &\multicolumn{2}{c}{2,409}  &\multicolumn{2}{c}{2,409} &\multicolumn{2}{c}{2,409}\\
$R^2$                  &\multicolumn{2}{c}{0.200} &\multicolumn{2}{c}{0.184}  &\multicolumn{2}{c}{0.164} &\multicolumn{2}{c}{0.190}\\
\bottomrule
\end{tabular}
\begin{tablenotes}
\footnotesize
    \item Data: \textit{kicker.de}; Bundesliga, first division, eight seasons from 2016/17 to 2023/24. Two ghost games before matchday 26 in the season 2019/20 are not included. Significance level: $^{*}p<0.10$, $^{**}p<0.05$, $^{***}p<0.01$. Heteroscedasticity-robust standard errors in parentheses.
    \end{tablenotes}
    \end{threeparttable}
    \end{table}

\begin{table}[htbp]\centering
\def\sym#1{\ifmmode^{#1}\else\(^{#1}\)\fi}
\caption{Without matchdays 18-25\label{table:18to25}}
\begin{threeparttable}
\fontsize{10}{16}\selectfont
\begin{tabular}{l *{4}{d{3}l}}
\toprule
           &\multicolumn{2}{c}{(1)}
           &\multicolumn{2}{c}{(2)}
          &\multicolumn{2}{c}{(3)}
           &\multicolumn{2}{c}{(4)}              \\
          &\multicolumn{2}{c}{\textbf{Goals}}
           &\multicolumn{2}{c}{\textbf{Goals}}
           &\multicolumn{2}{c}{\textbf{Win}}
           &\multicolumn{2}{c}{\textbf{Win}}         \\[-1ex]
 &\multicolumn{2}{c}{\textbf{home  team}}
           &\multicolumn{2}{c}{\textbf{away team}}
           &\multicolumn{2}{c}{\textbf{home team}}
           &\multicolumn{2}{c}{\textbf{away team}}         \\
\midrule
\multicolumn{9}{l}{\textbf{Season dummies, reference: season 18/19}}\\
\quad S16/17                             &  -0.102         &     (0.139)&      -0.173         &     (0.130)&       0.032         &     (0.049)&      -0.082\sym{*}  &     (0.044)      \\
\quad S17/18                             &  -0.062         &     (0.130)&      -0.184         &     (0.118)&       0.008         &     (0.048)&      -0.049         &     (0.043)      \\
\quad S19/20                             &  -0.007         &     (0.151)&       0.108         &     (0.138)&       0.004         &     (0.054)&       0.028         &     (0.051)      \\
 \quad S20/21                            &  -0.006         &     (0.135)&       0.050         &     (0.121)&      -0.046         &     (0.047)&       0.012         &     (0.044)      \\
\quad S21/22                             &  -0.155         &     (0.190)&       0.089         &     (0.172)&      -0.043         &     (0.066)&       0.019         &     (0.064)      \\
\quad S22/23                             &   0.110         &     (0.139)&       0.032         &     (0.122)&      -0.008         &     (0.047)&      -0.006         &     (0.043)      \\
\quad S23/24                             &   0.072         &     (0.145)&       0.062         &     (0.122)&      -0.011         &     (0.050)&      -0.011         &     (0.045)      \\
\multicolumn{9}{l}{\textbf{S19/20 $\times$ matchday 26-34} } \\
                                         &  -0.412\sym{**} &     (0.189)&       0.116         &     (0.187)&      -0.158\sym{**} &     (0.065)&       0.118\sym{*}  &     (0.064)     \\
\multicolumn{9}{l}{\textbf{S21/22 $\times$ stadium utilisation}}\\
\quad \qquad \quad \quad $[0,0.25]$        &   0.113         &     (0.312)&       0.638\sym{**} &     (0.316)&      -0.102         &     (0.110)&       0.120         &     (0.112)     \\
\quad \qquad \quad \quad  $(0.25,0.75]$   &   0.315         &     (0.193)&      -0.200         &     (0.171)&       0.108         &     (0.070)&      -0.067         &     (0.066)     \\
\textbf{Constant}                        &   1.767\sym{***}&     (0.102)&       1.349\sym{***}&     (0.088)&       0.461\sym{***}&     (0.034)&       0.305\sym{***}&     (0.031)     \\[1ex]
\midrule
\textbf{Fixed Effects} \\
\quad Home team           &\multicolumn{2}{c}{\checkmark}&\multicolumn{2}{c}{\checkmark}&\multicolumn{2}{c}{\checkmark}&\multicolumn{2}{c}{\checkmark}  \\
\quad Away team           &\multicolumn{2}{c}{\checkmark}&\multicolumn{2}{c}{\checkmark}&\multicolumn{2}{c}{\checkmark}&\multicolumn{2}{c}{\checkmark}  \\
\quad Matchday            &\multicolumn{2}{c}{\checkmark}&\multicolumn{2}{c}{\checkmark}&\multicolumn{2}{c}{\checkmark}&\multicolumn{2}{c}{\checkmark}  \\
\quad Ranking difference       &\multicolumn{2}{c}{\checkmark}&\multicolumn{2}{c}{\checkmark}&\multicolumn{2}{c}{\checkmark}&\multicolumn{2}{c}{\checkmark}  \\[-1ex]
\quad (home $-$ away team) \\
\midrule
$R^2$                  &\multicolumn{2}{c}{0.210} &\multicolumn{2}{c}{0.175}  &\multicolumn{2}{c}{0.176} &\multicolumn{2}{c}{0.189}\\
\bottomrule
\end{tabular}
\begin{tablenotes}
\footnotesize
    \item Data: \textit{kicker.de}; Bundesliga, first division, eight seasons from 2016/17 to 2023/24. Two ghost games before matchday 26 in the season 2019/20 are not included. Significance level: $^{*}p<0.10$, $^{**}p<0.05$, $^{***}p<0.01$. Heteroscedasticity-robust standard errors in parentheses.
    \end{tablenotes}
    \end{threeparttable}
    \end{table}

    \begin{table}[htbp]\centering
\def\sym#1{\ifmmode^{#1}\else\(^{#1}\)\fi}
\caption{Last nine matches of first and second half of seasons\label{table:lastnine}}
\begin{threeparttable}
\fontsize{10}{16}\selectfont
\begin{tabular}{l *{4}{d{3}l}}
\toprule
           &\multicolumn{2}{c}{(1)}
           &\multicolumn{2}{c}{(2)}
          &\multicolumn{2}{c}{(3)}
           &\multicolumn{2}{c}{(4)}              \\
          &\multicolumn{2}{c}{\textbf{Goals}}
           &\multicolumn{2}{c}{\textbf{Goals}}
           &\multicolumn{2}{c}{\textbf{Win}}
           &\multicolumn{2}{c}{\textbf{Win}}         \\[-1ex]
 &\multicolumn{2}{c}{\textbf{home  team}}
           &\multicolumn{2}{c}{\textbf{away team}}
           &\multicolumn{2}{c}{\textbf{home team}}
           &\multicolumn{2}{c}{\textbf{away team}}         \\
\midrule
\multicolumn{9}{l}{\textbf{Season dummies, reference: season 18/19}}\\
\quad S16/17                             &    -0.169         &     (0.176)&      -0.145         &     (0.160)&       0.008         &     (0.059)&      -0.077         &     (0.053)     \\
\quad S17/18                             &    -0.086         &     (0.165)&      -0.126         &     (0.151)&       0.004         &     (0.058)&      -0.050         &     (0.052)     \\
\quad S19/20                             &     0.074         &     (0.206)&       0.086         &     (0.193)&       0.035         &     (0.071)&       0.031         &     (0.066)     \\
 \quad S20/21                            &    -0.067         &     (0.170)&      -0.024         &     (0.146)&      -0.006         &     (0.058)&      -0.011         &     (0.054)     \\
\quad S21/22                             &    -0.184         &     (0.218)&       0.067         &     (0.187)&      -0.041         &     (0.073)&       0.010         &     (0.070)     \\
\quad S22/23                             &     0.229         &     (0.176)&      -0.017         &     (0.146)&       0.023         &     (0.057)&      -0.040         &     (0.051)     \\
\quad S23/24                             &     0.027         &     (0.182)&      -0.023         &     (0.150)&      -0.009         &     (0.061)&      -0.049         &     (0.054)     \\
\multicolumn{9}{l}{\textbf{S19/20 $\times$ matchday 26-34} } \\
&    -0.481\sym{**} &     (0.222)&       0.144         &     (0.221)&      -0.177\sym{**} &     (0.077)&       0.125\sym{*}  &     (0.075)    \\
\multicolumn{9}{l}{\textbf{S21/22 $\times$ stadium utilisation}}\\
\quad \qquad \quad \quad $[0,0.25]$        &     0.208         &     (0.327)&       0.631\sym{*}  &     (0.333)&      -0.065         &     (0.118)&       0.071         &     (0.117)    \\
\quad \qquad \quad \quad  $(0.25, 0.75]$   &     0.412\sym{*}  &     (0.245)&      -0.250         &     (0.214)&       0.189\sym{**} &     (0.088)&      -0.075         &     (0.084)    \\
\textbf{Constant}                        &     1.807\sym{***}&     (0.131)&       1.360\sym{***}&     (0.109)&       0.462\sym{***}&     (0.041)&       0.313\sym{***}&     (0.038)    \\[1ex]
\midrule
\textbf{Fixed Effects} \\
\quad Home team           &\multicolumn{2}{c}{\checkmark}&\multicolumn{2}{c}{\checkmark}&\multicolumn{2}{c}{\checkmark}&\multicolumn{2}{c}{\checkmark}  \\
\quad Away team           &\multicolumn{2}{c}{\checkmark}&\multicolumn{2}{c}{\checkmark}&\multicolumn{2}{c}{\checkmark}&\multicolumn{2}{c}{\checkmark}  \\
\quad Matchday            &\multicolumn{2}{c}{\checkmark}&\multicolumn{2}{c}{\checkmark}&\multicolumn{2}{c}{\checkmark}&\multicolumn{2}{c}{\checkmark}  \\
\quad Ranking difference       &\multicolumn{2}{c}{\checkmark}&\multicolumn{2}{c}{\checkmark}&\multicolumn{2}{c}{\checkmark}&\multicolumn{2}{c}{\checkmark}  \\[-1ex]
\quad (home $-$ away team) \\
\midrule
$R^2$                  &\multicolumn{2}{c}{0.205} &\multicolumn{2}{c}{0.177}  &\multicolumn{2}{c}{0.180} &\multicolumn{2}{c}{0.201}\\
\bottomrule
\end{tabular}
\begin{tablenotes}
\footnotesize
    \item Data: \textit{kicker.de}; Bundesliga, first division, eight seasons from 2016/17 to 2023/24. Two ghost games before matchday 26 in the season 2019/20 are not included. Significance level: $^{*} p < 0.10$, $^{**} p < 0.05$, $^{***} p < 0.01$.  Heteroscedasticity-robust standard errors in parentheses.
    \end{tablenotes}
    \end{threeparttable}
    \end{table}

\newgeometry{left=10mm, right=10mm, top=10mm, bottom=20mm}

    \begin{table}[htbp]\centering
\def\sym#1{\ifmmode^{#1}\else\(^{#1}\)\fi}
\caption{Post-match 25 dummy interacted with all seasons\label{table:postinteract}}
\begin{threeparttable}
\fontsize{10}{16}\selectfont
\begin{tabular}{l *{4}{d{3}l}}
\toprule
           &\multicolumn{2}{c}{(1)}
           &\multicolumn{2}{c}{(2)}
          &\multicolumn{2}{c}{(3)}
           &\multicolumn{2}{c}{(4)}              \\
          &\multicolumn{2}{c}{\textbf{Goals}}
           &\multicolumn{2}{c}{\textbf{Goals}}
           &\multicolumn{2}{c}{\textbf{Win}}
           &\multicolumn{2}{c}{\textbf{Win}}         \\[-1ex]
 &\multicolumn{2}{c}{\textbf{home  team}}
           &\multicolumn{2}{c}{\textbf{away team}}
           &\multicolumn{2}{c}{\textbf{home team}}
           &\multicolumn{2}{c}{\textbf{away team}}         \\
\midrule
\multicolumn{9}{l}{\textbf{Season dummies, reference: season 18/19}}\\
\quad S16/17                                     &    -0.052         &     (0.129)&      -0.347\sym{***}&     (0.115)&       0.074         &     (0.049)&      -0.077\sym{*}  &     (0.043)   \\       
\quad S17/18                                     &    -0.178         &     (0.119)&      -0.238\sym{**} &     (0.110)&      -0.011         &     (0.048)&      -0.050         &     (0.042)   \\       
\quad S19/20                                     &     0.090         &     (0.131)&       0.038         &     (0.122)&       0.013         &     (0.048)&       0.017         &     (0.043)   \\       
\quad S20/21                                     &    -0.000         &     (0.127)&       0.013         &     (0.116)&      -0.037         &     (0.047)&       0.008         &     (0.043)   \\       
\multicolumn{9}{l}{\quad \textbf{S21/22 $\times$ stadium utilisation}}\\                                          
\quad \qquad  $[0,0.25]$  \, \quad \quad (72 obs.)           &     0.070         &     (0.182)&       0.325\sym{*}  &     (0.184)&      -0.030         &     (0.067)&       0.063         &     (0.064)    \\      
\quad \qquad   $(0.25,0.75]$ \quad (129 obs.)           &     0.155         &     (0.153)&      -0.188         &     (0.135)&       0.060         &     (0.057)&      -0.057         &     (0.051)    \\      
\quad \qquad $(0.75,1.00]$  \quad (24 obs.)   &     0.071         &     (0.225)&      -0.236         &     (0.230)&       0.168\sym{*}  &     (0.098)&      -0.145\sym{*}  &     (0.080)    \\[1ex]      
\quad S22/23                                     &     0.191         &     (0.133)&      -0.086         &     (0.118)&       0.038         &     (0.048)&      -0.037         &     (0.043)    \\      
\quad S23/24                                     &     0.150         &     (0.134)&      -0.048         &     (0.116)&       0.008         &     (0.049)&      -0.039         &     (0.044)    \\      
\multicolumn{9}{l}{\textbf{Post matchday 26-34 dummies}} \\           
\quad  Post                                      &     0.342\sym{*}  &     (0.207)&      -0.027         &     (0.161)&       0.050         &     (0.059)&      -0.020         &     (0.056)   \\       
\quad Post $\times$ S16/17                       &    -0.094         &     (0.263)&       0.346         &     (0.236)&      -0.060         &     (0.087)&      -0.009         &     (0.078)   \\       
\quad Post $\times$ S17/18                       &     0.037         &     (0.262)&       0.070         &     (0.229)&       0.043         &     (0.088)&       0.014         &     (0.080)   \\       
\quad Post $\times$ S19/20                       &    -0.613\sym{**} &     (0.267)&       0.173         &     (0.228)&      -0.152\sym{*}  &     (0.082)&       0.112         &     (0.079)   \\       
\quad Post $\times$ S20/21                       &    -0.062         &     (0.264)&      -0.145         &     (0.218)&       0.084         &     (0.087)&      -0.057         &     (0.079)   \\       
\multicolumn{9}{l}{\quad \textbf{Post $\times$ S21/22 $\times$  stadium utilisation}}\\                        
\quad \qquad  $[0,0.25]$  \quad \quad \, (0 obs.)              \\      
\quad \qquad   $(0.25,0.75]$  \quad (19 obs.)          &    -0.161         &     (0.403)&       0.049         &     (0.375)&       0.048         &     (0.136)&      -0.003         &     (0.112)    \\      
\quad \qquad   $(0.75,1.00]$  \quad (62 obs.)         &    -0.462         &     (0.334)&       0.412         &     (0.303)&      -0.291\sym{**} &     (0.123)&       0.209\sym{*}  &     (0.109)    \\[1ex]      
\quad Post $\times$ S22/23                       &    -0.361         &     (0.270)&       0.052         &     (0.220)&      -0.071         &     (0.086)&       0.018         &     (0.078)    \\      
\quad Post $\times$ S23/24                       &    -0.422         &     (0.272)&       0.224         &     (0.223)&      -0.077         &     (0.086)&       0.053         &     (0.078)    \\      
\textbf{Constant}                                         &     1.661\sym{***}&     (0.091)&       1.413\sym{***}&     (0.084)&       0.432\sym{***}&     (0.034)&       0.327\sym{***}&     (0.030)    \\[1ex] 
\midrule
\textbf{Fixed Effects} \\
\quad Home team           &\multicolumn{2}{c}{\checkmark}&\multicolumn{2}{c}{\checkmark}&\multicolumn{2}{c}{\checkmark}&\multicolumn{2}{c}{\checkmark}  \\
\quad Away team           &\multicolumn{2}{c}{\checkmark}&\multicolumn{2}{c}{\checkmark}&\multicolumn{2}{c}{\checkmark}&\multicolumn{2}{c}{\checkmark}  \\
\quad Ranking difference       &\multicolumn{2}{c}{\checkmark}&\multicolumn{2}{c}{\checkmark}&\multicolumn{2}{c}{\checkmark}&\multicolumn{2}{c}{\checkmark}  \\[-1ex]
\quad (home $-$ away team) \\
\midrule
$R^2$                  &\multicolumn{2}{c}{0.190} &\multicolumn{2}{c}{0.166}  &\multicolumn{2}{c}{0.149} &\multicolumn{2}{c}{0.179}\\
\bottomrule
\end{tabular}
\begin{tablenotes}
\footnotesize
    \item Data: \textit{kicker.de}; Bundesliga, first division, eight seasons from 2016/17 to 2023/24. Two ghost games before matchday 26 in the season 2019/20 are not included. Significance level: $^{*}p<0.10$, $^{**}p<0.05$, $^{***}p<0.01$. Heteroscedasticity-robust standard errors in parentheses.
    \end{tablenotes}
    \end{threeparttable}
    \end{table}

\restoregeometry

\begin{figure}[ht!]\centering
\caption{Leave-out regressions on the probability of the home team winning}
\label{fig:leave_out}
\includegraphics[width=1.0\textwidth]{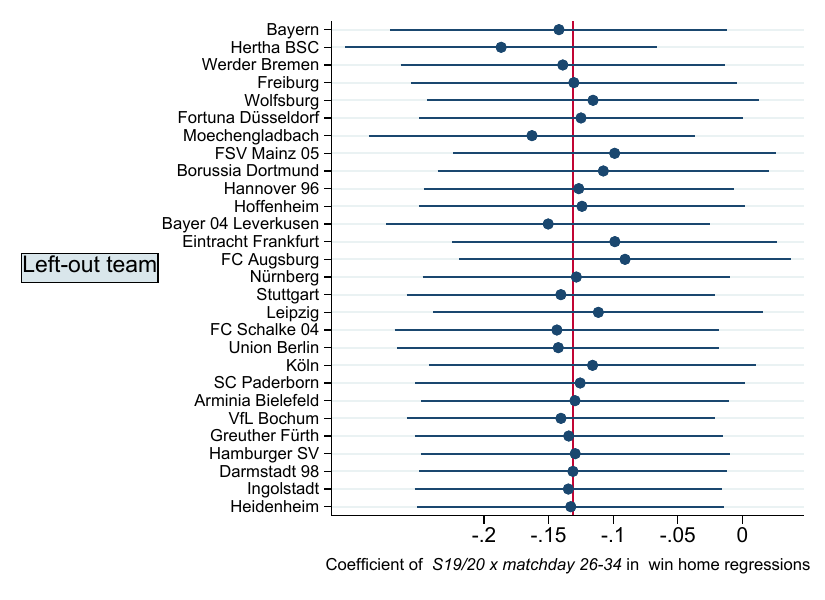}
\subcaption*{The figure shows the coefficient estimates of a dummy variable indicating the last nine matchdays of the 2019/20 season in a regression as specified in Equation~(\ref{eq:baseline}), with the dependent variable being equal to one if the home team won (and zero otherwise).  Each point represents the coefficient estimate of a different regression, with the left-out team indicated on the $y$-axis. The horizontal lines denote 95\% confidence intervals. The red vertical line indicates the value of the coefficient estimate when no team was left out ($-0.131$, see Table~\ref{table:diff_outcomes}, specification~(3)).}
\end{figure}

\begin{figure}[ht!]
\caption{\textbf{Away team} heterogeneity of first spectator ban period on home team winning probability}
\label{fig:het_away_team_home_win}
\includegraphics[width=0.7 \textwidth]{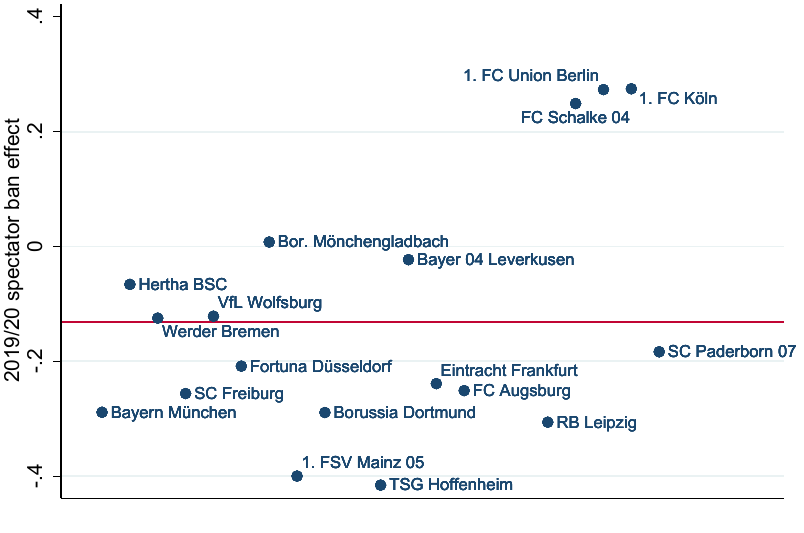}
\subcaption*{\footnotesize Data: \textit{kicker.de}, Bundesliga, first division, eight seasons from 2016/17 to 2023/24. Two ghost games in the season 2019/20 before matchday 26 are not included. 
Each point represents for matchdays $26-34$ of the season 2019/20 the (per \textbf{away team}) average difference between the actual value of the home win indicator and the predicted value of the home win indicator, where the predicted values are obtained from estimating Equation~(\ref{eq:baseline}) without  \texttt{S19/20 $\times$ matchday 26-34} for the sample without matchdays $26-34$ of the season 2019/20. The red horizontal line indicates the weighted average effect of $-0.132$. Note that this number marginally deviates from the reported coefficient on \texttt{S19/20 $\times$ matchday 26-34} in Table~\ref{table:diff_outcomes}, specification~(3), since for the latter the fixed effects estimates were identified from the full sample.
}
\end{figure}

\begin{figure}[ht!]
    \centering
 \caption{Effect of first spectator ban by ranking, stadium capacity and  utilisation}
    \label{fig:het_combined}
    \begin{subfigure}[t]{\textwidth}
        \centering
        \caption{Effect on goals home team}
        \includegraphics[width=0.9\textwidth]{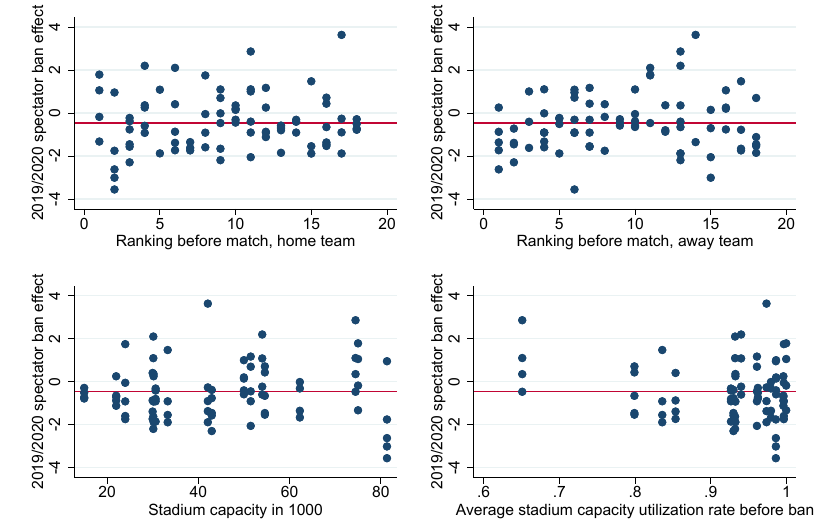}
                \label{fig:het_gaols_home_by}
    \end{subfigure}
    \vspace{0.5cm} 
    \begin{subfigure}[t]{\textwidth}
        \centering
        \caption{Effect on win home team}
        \includegraphics[width=0.9\textwidth]{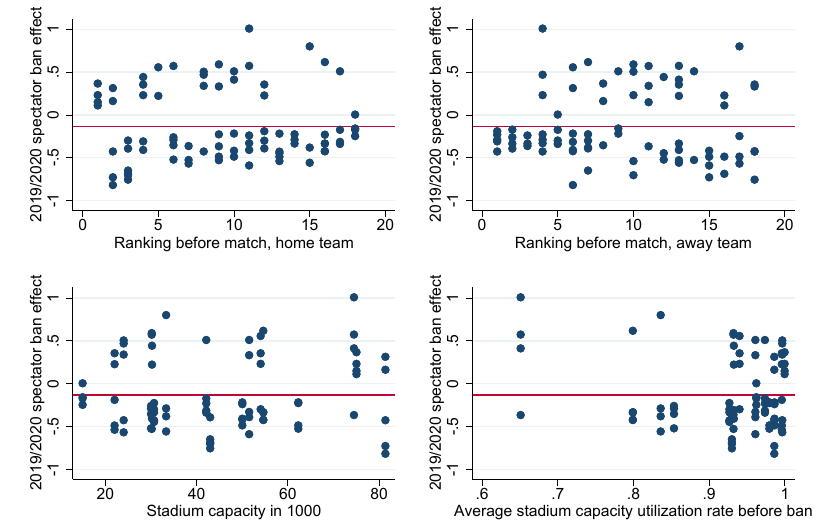}
        \label{fig:het_win_home_by}
    \end{subfigure}
 \vspace{-0.1cm}
 \subcaption*{\footnotesize Data: \textit{kicker.de}, Bundesliga, first division, eight seasons from 2016/17 to 2023/24. Two ghost games in the season 2019/20 before matchday 26 are not included. 
Each point represents for matchdays $26-34$ of the season 2019/20 the average difference between the actual value of the home goals  and the predicted value of the home goals (and analogously for home win), where the predicted values are obtained from estimating Equation~(\ref{eq:baseline}) without  \texttt{S19/20 $\times$ matchday 26-34} for the sample without matchdays $26-34$ of the season 2019/20. The red horizontal lines indicate the weighted average effect of home goals ($-0.469$)  respectively of a home win ($-0.132$). Note that these numbers marginally deviate from the reported coefficients on \texttt{S19/20 $\times$ matchday 26-34} in Table~\ref{table:diff_outcomes}, specifications~(1) and~(3), since for the latter the fixed effects estimates were identified from the full sample.
}
 \end{figure}

\end{document}